\documentclass[superscriptaddress,nobibnotes,amsmath,amssymb,notitlepage,twocolumn,pra,longbibliography]{revtex4-2}

\usepackage{bm,braket}
\usepackage[toc,page]{appendix}
\usepackage{comment}
\usepackage{dcolumn}

\usepackage{amsfonts}

\usepackage{graphicx,color,hyperref}
\usepackage[caption=false]{subfig}
\hypersetup{colorlinks=true, linkcolor=blue, citecolor=blue, urlcolor=blue} 

\graphicspath{{./img/}}

\newcommand{\beq}[1]{\begin{equation}\label{#1}}
\newcommand{\eep}{\;.\end{equation}}
\newcommand{\eec}{\;,\end{equation}}
\newcommand{\eeq}{\end{equation}}

\newcommand*\dd{\mathop{}\!\mathrm{d}} %differential d

\newcommand{\om}{\omega}

\DeclareMathAlphabet{\mathcal}{OMS}{cmsy}{m}{n} % Changes font for mathcal but leaves the rest of the math fonts in Times.

\renewcommand{\vec}[1]{{\bf #1}}

\newcommand{\kv}{\vec{k}}

\newcommand{\A}{\vec{A}}

\usepackage{amsmath}
\usepackage{amssymb}
\usepackage{xcolor}
\usepackage{bbm}
\usepackage{physics}
\usepackage{float}
\usepackage{graphicx}
\usepackage{dcolumn} % Align table columns on decimal point
\usepackage{bm} % bold math
\usepackage{siunitx}
\usepackage{enumitem}  % to use (i) etc for enumerate lists
\makeatletter
\renewcommand*{\fnum@figure}{{\normalfont\bfseries \figurename~\thefigure}}
\makeatother

\allowdisplaybreaks

\definecolor{orange}{rgb}{1,0.5,0}

\newcommand{\sect}[1]{\vspace{0.3em}{\it #1.}---}

\graphicspath{{./img/}}

\DeclareMathAlphabet{\mathcal}{OMS}{cmsy}{m}{n} % Changes font for mathcal but leaves the rest of the math fonts in Times.

\newcommand{\ii}{\mathrm{i}}

\renewcommand{\Re}{\mathfrak{Re}~}

\makeatletter
\newcommand{\specificthanks}[1]{\@fnsymbol{#1}}% Inserts a specific \thanks symbol
\makeatother

\begin{document}

\preprint{APS/123-QED}

\title{Optical manifestations and bounds of topological Euler class}

\author{Wojciech J. Jankowski}
\email{wjj25@cam.ac.uk}
\thanks{}
\affiliation{TCM Group, Cavendish Laboratory, Department of Physics, J J Thomson Avenue, Cambridge CB3 0HE, United Kingdom}

\author{Arthur S. Morris}
\thanks{}
\affiliation{TCM Group, Cavendish Laboratory, Department of Physics, J J Thomson Avenue, Cambridge CB3 0HE, United Kingdom}

\author{Adrien Bouhon}
\thanks{}
\affiliation{TCM Group, Cavendish Laboratory, Department of Physics, J J Thomson Avenue, Cambridge CB3 0HE, United Kingdom}
\affiliation{Nordita, Stockholm University and KTH Royal Institute of Technology, Hannes Alfv{\'e}ns v{\"a}g 12, SE-106 91 Stockholm, Sweden}

 \author{F. Nur \"Unal}
\thanks{}
\affiliation{TCM Group, Cavendish Laboratory, Department of Physics, J J Thomson Avenue, Cambridge CB3 0HE, United Kingdom}
\affiliation{School of Physics and Astronomy, University of Birmingham, Edgbaston, Birmingham B15 2TT, United Kingdom}%\looseness=-1}

\author{Robert-Jan Slager}
\email{rjs269@cam.ac.uk}
\affiliation{TCM Group, Cavendish Laboratory, Department of Physics, J J Thomson Avenue, Cambridge CB3 0HE, United Kingdom}
\affiliation{Department of Physics and Astronomy, University of Manchester, Oxford Road, Manchester M13 9PL, United Kingdom}

\date{\today}

\begin{abstract}
We analyze quantum-geometric bounds on optical weights in topological phases with pairs of bands hosting nontrivial Euler class, a multigap invariant characterizing non-Abelian band topology. We show how the bounds constrain the combined optical weights of the Euler bands at different dopings and further restrict the size of the adjacent band gaps. In this process, we also consider the associated interband contributions to dc conductivities in the flat-band limit. We physically validate these results by recasting the bound in terms of transition rates associated with the optical absorption of light, and demonstrate how the Euler connections and curvatures can be determined through the use of momentum and frequency-resolved optical measurements, allowing for a direct measurement of this multiband invariant. Additionally, we prove that the bound holds beyond the degenerate limit of Euler bands, resulting in nodal topology captured by the patch Euler class. In this context, we deduce optical manifestations of Euler topology within $\vec{k} \cdot \vec{p}$ models, which include quantized optical conductivity, and third-order jerk photoconductivities. We showcase our findings with numerical validation in lattice-regularized models that benchmark effective theories for real materials and are realizable in metamaterials and optical lattices.
\end{abstract} 

\maketitle

\sect{Introduction} Topological insulators and semimetals constitute an active field~\cite{Rmp1,Rmp2,Armitage_2018}. Topological insulators give rise to a condensed--matter realization of the $\theta$ vacuum, exhibiting a parity anomaly~\cite{Qi_2008,PhysRevLett.61.2015}, resulting in a variety of physical phenomena~\cite{de_Juan_2017, doi:10.1126/sciadv.1701207, PhysRevB.97.201117, PhysRevB.99.161404, Parker_2019, PhysRevLett.122.210401, jacobrjs, Asteria_2019, Avdoshkin_2020, Gianfrate2020}. Motivated by the recent progress in the understanding of such relationships between topology and optical responses, and further by the experimental observation of effects that result from this connection, a natural question is to what extent these insights translate to the context of recently discovered multigap topologies with non-Abelian properties. 

While conventional ``single-gap'' topologies~\cite{Kitaevtenfold, SchnyderClass, Slager_NatPhys_2013, Clas1, Ryu_2010, fukane, prx_us,BernevigHughes, rjs_translational, Shiozaki14,PhysRevB.100.195135, Clas4, Clas5,Ft1} are well understood, and can be classified by comparing general momentum-space constraints~\cite{prx_us} and real-space conditions~\cite{Clas4, Clas5}, a variety of problems concerning multigap topologies~\cite{Bouhon_2020_geo, davoyan2023mathcalpmathcaltsymmetric} remain open. A prominent example in this regard is the Euler class, which characterizes systems with real Hamiltonians (the reality condition is usually guaranteed by the presence of either $\mathcal{C}_2\mathcal{T}$ or $\mathcal{PT}$ symmetry). In such systems, band degeneracies residing between adjacent pairs of bands can carry non-Abelian frame charges~\cite{doi:10.1126/science.aau8740, PhysRevX.9.021013, Bouhon_2019}, akin to $\pi$-disclination defects in biaxial nematics~\cite{Kamienrmp, Genqcs2016,volovik2018investigation, Beekman20171}, which can be changed upon braiding these nodes in momentum space. In particular, two-band subspaces may feature a finite Euler class that physically manifests itself by having pairs of nodes with similarly valued frame charges that cannot annihilate each other. The Euler class of a given two-band subspace can be evaluated on any patch $\mathcal{D}$ in the Brillouin zone (BZ) as~\cite{Bouhon_2019, jankowski2023disorderinduced}
\begin{equation}
  \chi = \frac{1}{2\pi} \int_{\mathcal{D} \subseteq\text{BZ}} \dd ^{2} \textbf{k}~\text{Eu}(\kv) - \frac{1}{2\pi} \oint_{\partial\mathcal{D}} \dd \textbf{k} \cdot \vec{a}(\kv) .
  \label{eq:eulerpatch}
\end{equation}
Here $\vec{a}$ denotes the Euler connection, defined as the Pfaffian of the non-Abelian Berry connection $\A_{n,n+1}(\kv) \equiv \bra{u_n(\kv)}\ket{\nabla_{\kv} u_{n+1}(\kv)}$ of the neighboring Bloch bands $\ket{u_n(\kv)}$ and $\ket{u_{n+1}(\kv)}$, while $\text{Eu}(\kv) = \nabla_{\textbf{k}} \cross \vec{a}(\kv)$ is the non-Abelian Euler curvature, equal to the exterior derivative of the Euler connection with respect to the quasimomentum $\kv$~\cite{Bouhon_2019}. In particular, when the patch covers the entire Brillouin zone, the boundary $\partial \mathcal{D}$ vanishes and Eq.~\eqref{eq:eulerpatch} acts as a real analog of the Chern number, 
\begin{equation}\label{eq:chern}
 C = \frac{1}{2\pi} \int_{\text{BZ}} \dd ^{2} \textbf{k} ~ \mathcal{F}(\kv),
\end{equation}
which is, instead, given by the Abelian Berry curvature ${\mathcal{F}(\kv) =\ii\sum^\text{occ}_{n}\nabla_\kv \times \vec{A}_{n,n}}$, with $\vec{A}_{n,n}$ the single-band connection.  We note that manifestations of multigap topologies, the braiding of non-Abelian charges, and the Euler class have been found in a variety of physical settings and phenomena, including quench dynamics~\cite{Unal_2020,Zhao_2022}, periodically driven Floquet systems~\cite{rj_floquet}, and contexts that range from phononic~\cite{park2021,Lange2022,Peng2021,peng2022multi} and electronic systems~\cite{chen2021manipulation, Bouhon_2019}, to twisted bilayer and magnetic systems \cite{magnetic,PhysRevX.9.021013}, or acoustic and photonic metamaterials~\cite{Jiang_2021, Guo1Dexp}.

Motivated by a range of works reporting an interplay between Chern numbers and light-matter responses \cite{PhysRevLett.103.116803, doi:10.1126/sciadv.1701207, Asteria_2019}, we here address the question of whether optical observables can capture the topological Euler class. Intuitively, these relationships arise from quantum-geometric constraints imposed by the invariant; quantum geometry is directly related to optical properties~\cite{Ahn_2021, bouhon2023quantum}. Using a recent formulation~\cite{bouhon2023quantum} of the many-band quantum metric using Pl\"ucker maps, we derive three analytical bounds on: (i)~optical weights, detailing relations of interband contributions to dc conductivities induced by the quantum metric, (ii)~combined optical transition rates under circularly polarized light, and (iii)~gaps both above {\it and} below the Euler bands, providing a multigap extension of the single-gap bound on Chern insulators recently derived in Ref.~\cite{onishi2023fundamental}. We show that our results concerning optical weights hold in degenerate flat Euler bands, e.g.,~conjectured in the context of twisted bilayer graphene (TBG) \cite{PhysRevX.9.021013}, and beyond, i.e.,~when nodal Euler topology arises in the presence of dispersion~\cite{Bouhon_2019, Jiang_2021}. Accordingly, we discover optical manifestations of such nodes in terms of (i) quantized optical conductivities, (ii) optical weights, and (iii) third-order jerk photoconductivities~\cite{PhysRevLett.121.176604, PhysRevB.100.064301}, within effective $\vec{k} \cdot \vec{p}$ models. We further demonstrate how the associated Euler invariants can be reconstructed in terms of optical transition rates. Finally, we numerically showcase the obtained identities in  minimal lattice-regularized models, setting the stage for effective theories in real materials, whilst also offering direct implementations in ultracold atomic simulations and metamaterials. %We start by considering optical manifestations of isolated Euler bands.

\begin{figure}
\centering
%  \captionsetup[figure]{justification=justified, singlelinecheck=off} 
  \includegraphics[width=0.9\columnwidth]{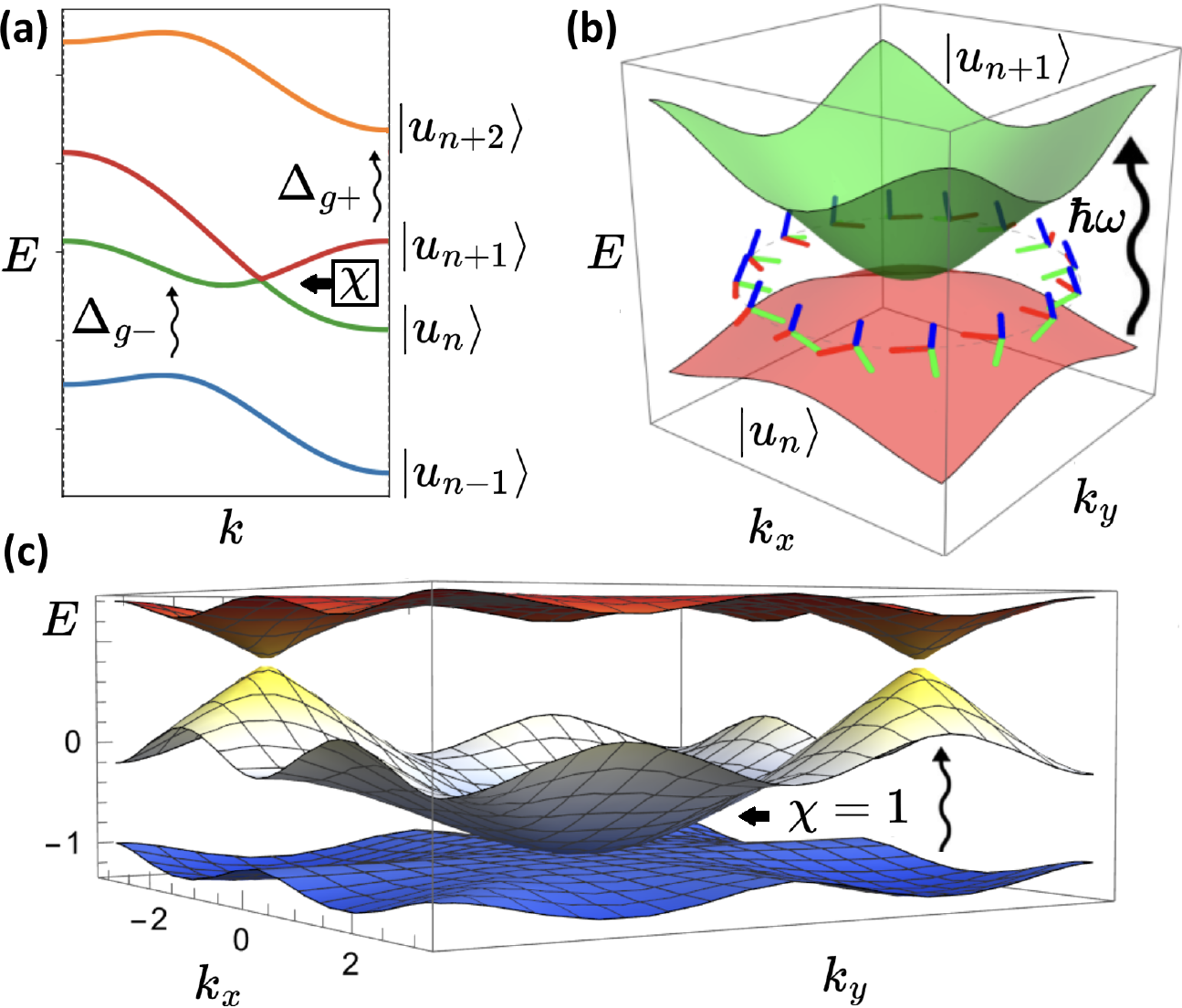}
  \caption{\small \textbf{(a)} Optical transitions to/from Euler bands with lower/upper gaps having energies $\Delta_{g-}$/$\Delta_{g+}$. \textbf{(b)} Quadratic Euler node with nontrivial patch Euler class $\chi = 1$ in a $\vec{k} \cdot \vec{p}$ model. The associated eigenstate frame winding (colored vectors) corresponds to the winding of the non-Abelian connection $\textbf{A}_{1,2}$, which is singular at the node. An optical transition within the Euler band subspace (for semimetallic doping) is schematically shown. \textbf{(c)} Lattice-regularized realization of quadratic Euler node in a kagome model. The optical conductivities, optical weights, as well as jerk conductivities, can be used to deduce the patch Euler class~$\chi$.}
\label{fig:effective}
\end{figure}

\sect{Optical manifestations of Euler bands}
We first consider the case of isolated Euler bands and discuss bounds on the optical observables captured by quantum geometric quantities, which can be indeed used to infer the presence of a nontrivial Euler class. We label the magnitude of the (direct) gaps below and above the Euler bands $\{ \ket{u_{n}},\ket{u_{n+1}} \}$ of energies $E_{n}(\kv)$, $E_{n+1}(\kv)$ as $\Delta_{g-} \equiv \text{min}_\kv \left[ E_{n} (\kv) - E_{n-1} (\kv) \right]$ and $\Delta_{g+} \equiv \text{min}_\kv \left[ E_{n+2} (\kv) - E_{n+1} (\kv) \right]$ respectively (see Fig.~\ref{fig:effective}(a)). Before deriving the bounds, we recall the inequality imposed on quantum geometry due to the nontriviality of the Euler class \cite{Ahn_2021, bouhon2023quantum}, as it forms the basis of further calculations. The quantum metric is defined as $g^\chi_{ab} (\kv) = \Re  \sum^{2}_{n=1} \sum_{m \neq 1,2} \left[r^a_{nm}r^b_{mn}\right]$, where $a,b = x,y$, and the ${r^a_{nm} = \ii(1-\delta_{n,m})A^a_{n,m}}$ are transition dipole matrix elements captured by the components of the non-Abelian Berry connection $A^a_{n,m}$. The integral of the trace of $g^\chi_{ab}$ over the BZ can be shown to be bounded from below by the Euler class of the two-band subspace, % (see Sec.~\ref{app:appA}) \cite{PhysRevLett.124.167002, bouhon2023quantum}:
\beq{eq:inequality}
    \frac{1}{4\pi} \int \dd^2 \vec{k} ~ \Tr\textbf{g}^\chi \geq |\chi|,
\eeq
where the trace is over $\vec{k}$-space coordinates at each $\vec{k}$-point individually (see Refs.~\cite{PhysRevLett.124.167002, bouhon2023quantum, PhysRevLett.131.156901} and Supplemental Material (SM)~\cite{SI}). On the other hand, in three-band models, the Euler class is captured by the winding of the third gapped vector $\textbf{n}_3\equiv \ket{u_3}$, and may be expressed as ${\chi = \frac{1}{4\pi} \int \dd^2 \vec{k} ~ \textbf{n}_3 \cdot (\partial_{k_x} \textbf{n}_3 \cross \partial_{k_y} \textbf{n}_3)}$~\cite{Unal_2020}. In this case, the left-hand side of Eq.~\eqref{eq:inequality} reduces to $\frac{1}{4\pi} \int \dd^2 \vec{k}  ~ \Tr\textbf{g}^\chi = \frac{1}{4\pi} \int \dd^2 \vec{k}  ~ {(|\partial_{k_x} \textbf{n}_3|^2 + |\partial_{k_y} \textbf{n}_3|^2)}$. In fact, this is the result of a stronger inequality that holds at every point in the BZ (i.e.,~between the integrands). Namely, ${|\partial_{k_x} \textbf{n}_3|^2 + |\partial_{k_y} \textbf{n}_3|^2} \geq {|\textbf{n}_3 \cdot (\partial_{k_x} \textbf{n}_3 \cross \partial_{k_y} \textbf{n}_3)|}$, which is analogous to a similar geometric bound in Chern insulators, which arises from the skyrmion formula~\cite{PhysRevB.94.134423, onishi2023fundamental}. This inequality suggests an interplay between the Euler class and optical phenomena. 

In that context, the quantum metric is related to the intrinsic quadrupole moment, the quantity upon which electric quadrupole (E2) transitions depend~\cite{Ahn_2021, Pozo_Oca_a_2023}. In Euler phases, since the diagonal elements of the Berry connection vanish due to the reality condition, this culminates in the single-particle quadrupole moment given by ${q^n_{ab}(\kv) \equiv -\frac{e}{2}\left(\bra{\partial_{k_a}u_{n}}\ket{\partial_{k_b}u_{n}} + \bra{\partial_{k_b}u_{n}}\ket{\partial_{k_a}u_{n}}\right)}$, where $e$ is the elementary charge. Consequently, by employing $q^1_{ab} + q^2_{ab} = -eg^{\chi}_{ab}$ for a real Hamiltonian, we find that Eq.~\eqref{eq:inequality} leads to a bound on the integrated total quadrupole moment of the topological bands due to their finite Euler class,
\beq{eq:quadrupole}
    \frac{1}{4\pi} \int \dd^2 \vec{k} ~|\Tr (q^1_{ab} + q^2_{ab})| \geq e|\chi|.
\eeq
Here, we consider coupling to an electric field ${\mathcal{E} = (\mathcal{E}_x \cos \om t, \mathcal{E}_y \cos(\om t + \phi))}$, where the phase shift $\phi$ controls the polarization, with $\phi = \pm\pi/2$ corresponding to left- and right-circularly polarized light (LCP/RCP). In this context, the quantum metric can be accessed by inspection of the appropriate optical weight ${W^1_{ab}(\om_{\text{max}}) \equiv \int^{\om_{\text{max}}}_0 \dd \om~ \frac{\Re  \sigma_{ab}(\om)}{\om}}$, for the  optical conductivities $\sigma_{ab}(\om)$ under the electric field up to some large frequency $\om_{\text{max}}$~\cite{PhysRevB.97.201117,Gianfrate2020,Tan19_PRL_QGT_SCqubit,yu2022experimental,SI}. % (see Sec.~II of SM~\cite{SI} for further elaboration).

\begin{figure}
\centering
%  \captionsetup[figure]{justification=justified, singlelinecheck=off} 
  \includegraphics[width=1.0\columnwidth]{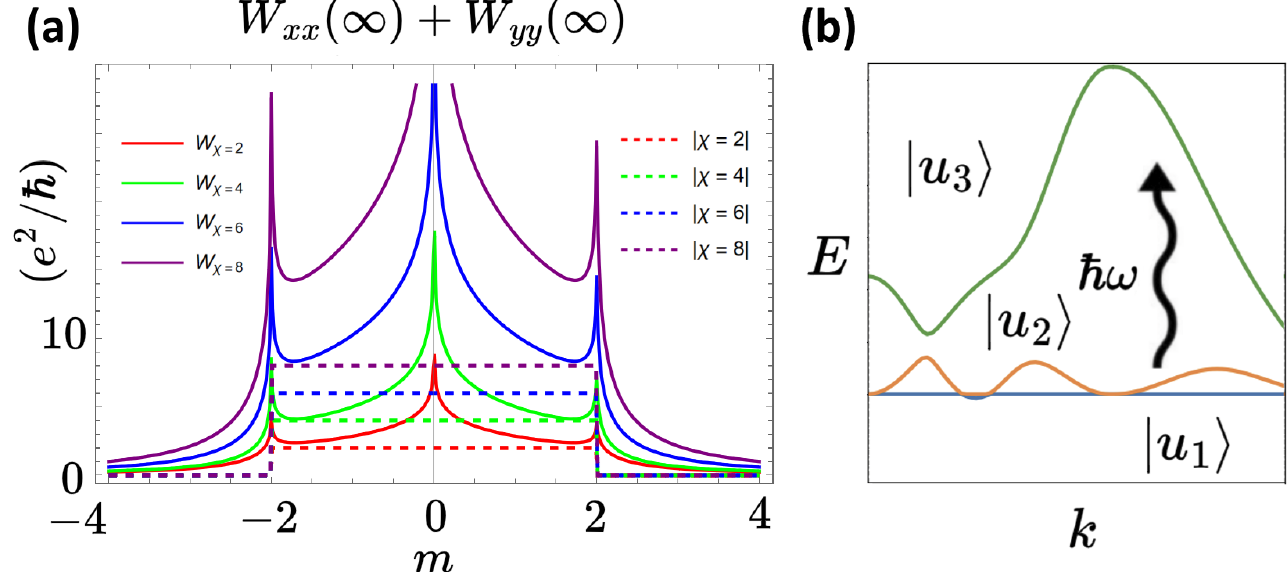}
  \caption{ \textbf{(a)} Optical weights in a three-band Euler Hamiltonian with the occupied Euler bands [lowest two bands as depicted in \textbf{(b)}] can be related to the interband contributions to linear dc conductivities. The combined optical weight is bounded by the Euler invariant, see Eq.~\eqref{eq:W_bound_3band}.
  Singularities occur at topological phase transitions ($|m|=2$), associated with the closing of the gap as the topological mass term $m$ is changed. The limit of large $|m|$ corresponds to the trivial phase with $\chi = 0$. \textbf{(b)} A schematic picture of the corresponding transitions.}
\label{fig:Ws}
\end{figure}

We now focus on the optical weights associated with transitions to and from the Euler subspace, which apply to arbitrarily dispersive Euler bands such as those arising in lattice models (for examples see SM~\cite{SI}, Refs.~\cite{bouhon2022multigap, rj_floquet, Jiang_2021}). 
As our first central result, we 
derive an optical bound constraining the optical weights $W^1_{ab}$ involving the Euler bands, across the gaps harbouring $\Delta_{g-}$ and $\Delta_{g+}$, (see Fig.~\ref{fig:effective})~\cite{PhysRevB.62.1666,onishi2023fundamental}. We label optical transitions into the isolated Euler subspace from bands below, and out of this subspace into bands above by unoccupied (unocc) and occupied (occ), respectively. We find that these transitions involving the Euler bands are bounded as
\begin{align}
\label{eq:opticalBound}
\begin{split}
    &W^1_{xx,(\text{occ})}(\infty) + W^1_{yy,(\text{occ})}(\infty) \\ 
    &+ W^1_{xx,(\text{unocc})}(\infty) + W^1_{yy,(\text{unocc})}(\infty)  
    \geq \frac{e^2}{\hbar} |\chi|,
\end{split}
\end{align}
which holds in the high-frequency limit $\om_{\text{max}} \rightarrow \infty$~\cite{SI}.
%The corresponding transitions occur across gaps of energies $\Delta_{g-}$ and $\Delta_{g+}$ (see Fig.~\ref{fig:effective}). 
If there are no bands below the Euler subspace, ${W^1_{aa,(\text{unocc})}(\infty) }$ vanishes and this condition reduces to,
\beq{} \label{eq:W_bound_3band}
    W^1_{xx,(\text{occ})}(\infty) + W^1_{yy,(\text{occ})}(\infty)
    \geq \frac{e^2}{\hbar} |\chi|,
\eeq
which is observable for topological phases with $\chi \neq 0$. We numerically corroborate these results in models with different Euler class (see Fig.~\ref{fig:Ws}). Finally, we point out that we can generalize these results to quantum-metric signatures in flat-band systems with dc linear conductivity due to the interband contributions~\cite{PhysRevB.105.085154}, as detailed in the SM~\cite{SI}.

We remark that an analogous conclusion, viewed as a bound on the optical weight, can be also reached  from Fermi's golden rule under circular polarization~\cite{PhysRevB.97.201117}. Namely, we find that the combined absorption rate $ \tilde{\Gamma}^\text{tot}$ of LCP and RCP light~\cite{doi:10.1126/sciadv.1701207, Asteria_2019}, combining all possible transitions when Euler bands are fully occupied and are completely unoccupied, is bounded from below by the Euler class, 
\beq{}\label{eq:FGRbound}
    \tilde{\Gamma}^\text{tot} = \frac{2\pi e^2 \mathcal{E}^2 }{\hbar^2} \int \dd^2 \vec{k}~ \Tr\textbf{g}^{\chi}~ \geq \frac{e^2 \mathcal{E}^2 }{2 \hbar^2} |\chi|.
\eeq
This result is consistent with Eq.~\eqref{eq:opticalBound}, on recognizing that combinations of LCP and RCP can be viewed as linear polarizations inducing corresponding excitations that appear in the optical weights~\cite{SI}. 

We conclude by showing that the bound on the optical weight is additionally related to the size of the gaps. To prove this, we start by considering a three-band spectrum with the Euler bands at the bottom, and apply the $f$-sum rule~\cite{onishi2023fundamental}, to recognize that $\int^\infty_0 \dd \om~\frac{\Re  \sigma_{aa}}{\om} \leq \int^\infty_0 \dd \om~\Re  \sigma_{aa} \frac{\hbar}{\Delta_{g+}} = \frac{n_{e,\text{occ}} \hbar e^2}{2m\Delta_{g+}}$. We obtain that
\beq{eq;single-gap-bound}
    \frac{n_{e,\text{occ}} \pi \hbar^2}{m \Delta_{g+}} \geq |\chi| \implies  \frac{n_{e,\text{occ}} \pi\hbar^2} {m|\chi|} \geq \Delta_{g+},
\eeq
where $n_{e,\text{occ}}$ is the charge density in the occupied bands, and $m$ is the bare electron mass. A tighter bound can be obtained upon replacing $m$ (from the minimal-coupling Hamiltonian) with the effective mass $m^*$ in the $f$-sum rule (within the $\kv \cdot \vec{p}$ approach)~\cite{onishi2023fundamental}. In the flat-band limit, a non-zero Euler class $|\chi| \neq 0$ implies that the gap $\Delta_{g+}$ vanishes in the limit of large effective mass $m^* \rightarrow \infty$. Hence, to keep the gap finite, an Euler insulator needs dispersion in either the Euler bands or in the unoccupied band above the gap. In the many-band case, we analogously employ the sum rule to obtain,
\beq{}
    \frac{\pi \hbar^2}{m}\left(\frac{n_{e,\text{occ}}}{\Delta_{g-}} + \frac{n_{e,\text{unocc}}}{ \Delta_{g+}}\right) 
    \geq |\chi|,
\eeq
where $n_{e,\text{occ}}, n_{e,\text{unocc}} \sim 1/a^2$, with lattice constant $a$, are the total charge densities for the cases where the Euler bands are occupied (as before) and unoccupied, respectively. By demanding that $n_{e,\text{occ}}$ is the carrier density in certain $\kv \cdot \vec{p}$ low-energy models, e.g., for twisted bilayer graphene~\cite{PhysRevLett.122.106405, Bernevig_2021, bennett2023twisted}, a further bound can be obtained. Here, the Euler class constrains the size of {\it both} gaps as
\beq{eq:gapbound}
    \frac{n_{e,\text{unocc}} \pi \hbar^2}{m|\chi|}
    \geq \frac{n_{e,\text{unocc}}}{\frac{n_{e,\text{occ}}}{ \Delta_{g-}}+\frac{n_{e,\text{unocc}}}{ \Delta_{g+}}} > \frac{1}{2} \text{HM}(\Delta_{g-}, \Delta_{g+}),
\eeq
where $\text{HM}(\Delta_{g-}, \Delta_{g+}) \equiv [(\Delta_{g+}^{-1}+\Delta_{g-}^{-1})/2]^{-1}$ is the harmonic mean of the band gaps, and we have used $n_{e,\text{occ}} > n_{e,\text{unocc}}$. This is the central result of this work, elucidating a quantum-geometric constraint on the multigap
Euler topology in electronic materials, as manifested by a
fundamental optical bound on the band gaps. Contrary to the bounds related to dc conductivity within noninteracting particle picture (see SM~\cite{SI}), this kind of fundamental bound could even hold in interacting, strongly correlated systems~\cite{onishi2023fundamental}.

\sect{Optical manifestations of patch Euler class} 
We now demonstrate how our findings generalize for a patch Euler invariant, which extend beyond subspaces separated by gaps $\Delta_{g\pm}$ (see Fig.~\ref{fig:effective}). A patch of two bands with $\chi \neq 0$ necessarily hosts band nodes with same non-Abelian frame charges that are obstructed to annihilate~\cite{PhysRevB.102.115135}, as long as the patch excludes band crossings of other bands~\cite{Bouhon_2019}. However, accessing this invariant can be complicated due to the multiband character of the Euler class~\cite{PhysRevB.102.115135}.
We here show that it can be probed with optical conductivities, associated optical weights, and third-order jerk photoconductivities~\cite{PhysRevLett.121.176604, PhysRevB.100.064301}.

A general $\textbf{\kv} \cdot \vec{p}$ Hamiltonian for a bulk Euler node hosting a patch Euler class can be written as
\beq{eq:EulerNode}
H_\chi(\kv)=\alpha(k_+^{2|\chi|}\sigma_-+k^{2|\chi|}_-\sigma_+),
\eeq
where $\alpha$ is a dispersion constant, $k_{\pm}=k_x \pm \ii k_y$ and ${\sigma_{\pm}=(\sigma_x\pm\ii\sigma_z)/2}$~\cite{morris2023andreev}. Note that the Hamiltonian is real by construction. The elements of the non-Abelian Berry connection, the quantum metric, and associated optical responses, may be computed from the eigenstates of $H_\chi(\kv)$. We find that in the proximity of the node, the optical conductivity is quantized by the patch Euler invariant $\chi$ as
\beq{eq:quantizedACnodal}
\sigma_{aa}(\om) = \frac{e^2}{8\hbar}|\chi|.
\eeq
Furthermore, on doping with chemical potential $\mu$ away from the node (see SM~\cite{SI}), we show that the optical weight is
\beq{eq:nodalW}
    W^1_{aa}(\om_{\text{max}}) = \frac{\pi  e^2}{4h} |\chi| \text{ln} \left( \frac{\om_{\text{max}}}{2\mu} \right).
\eeq
Since $\om_\text{max}$ and $\mu$ are controllable parameters,  we predict that the patch Euler class can be not only deduced with optical conductivities, but also traced on doping.

As a next step, we demonstrate that independently of the dispersion proportionality constant $\alpha$, the jerk conductivities~\cite{PhysRevLett.121.176604} at any non-vanishing light frequency $\om$ offer universal Euler class-dependent ratios~\cite{SI}. Most interestingly, we obtain that
\beq{eq:nodal_jerk}
    \frac{\sigma^{xxxx}_{\text{jerk}}(\om)}{\sigma^{xxyy}_{\text{jerk}}(\om)} = \frac{3|\chi|-1}{|\chi|+1},
\eeq
which is always non-vanishing and bounded: ${1 <\sigma^{xxxx}_{\text{jerk}}/\sigma^{xxyy}_{\text{jerk}} < 3}$. This holds provided that, at least to first order, the node (i) is rotationally symmetric, and (ii) hosts patch Euler class $\chi$. We also note that the nature of the low-frequency divergence of jerk photoconductivities $\om \rightarrow 0$ depends on $\chi$, becoming weaker as $\chi$ increases (see SM~\cite{SI}). All second-order photoconductivities vanish within the model, as the bulk node naturally enjoys the inversion symmetry~\cite{SI}. When full rotational symmetry is not preserved, as in lattice models, we find that Eqs.~\eqref{eq:quantizedACnodal}--\eqref{eq:nodal_jerk} still hold closely to the nodes. 

\sect{Reconstructing the Euler invariant}
Finally, we demonstrate that the Euler invariant \eqref{eq:eulerpatch} can be probed by monitoring the transition rates in a frequency- and momentum-resolved way. Motivated by the reconstruction of the  quantum geometric tensor (QGT)~\cite{PhysRevB.97.201117,Gianfrate2020}, we devise a protocol upon considering \textit{only}  the bottom band of the Euler subspace to be occupied (not both), see~Fig.~\ref{fig:effective}(b). We recognize that both terms of Eq.~\eqref{eq:eulerpatch} can be extracted from the elements of the non-Abelian Berry connection, which we identify as real part of the QGT. Specifically, the vector components can be deduced from the absorption rate of the linearly polarized light in the $x$ and $y$ direction, probing $g^{12}_{xx} = |A^x_{12}|^2$, and $g^{12}_{yy} = |A^y_{12}|^2$, see SM~\cite{SI}. The non-Abelian connection vector can be thus obtained modulo the sign of each component. Imposing a smooth real gauge by hand to fix the sign as a gauge choice, we then compute the Euler curvature and the Euler invariant.

\begin{figure}[t]
%  \captionsetup[figure]{justification=justified, singlelinecheck=off} 
  \includegraphics[width=1.0\columnwidth]{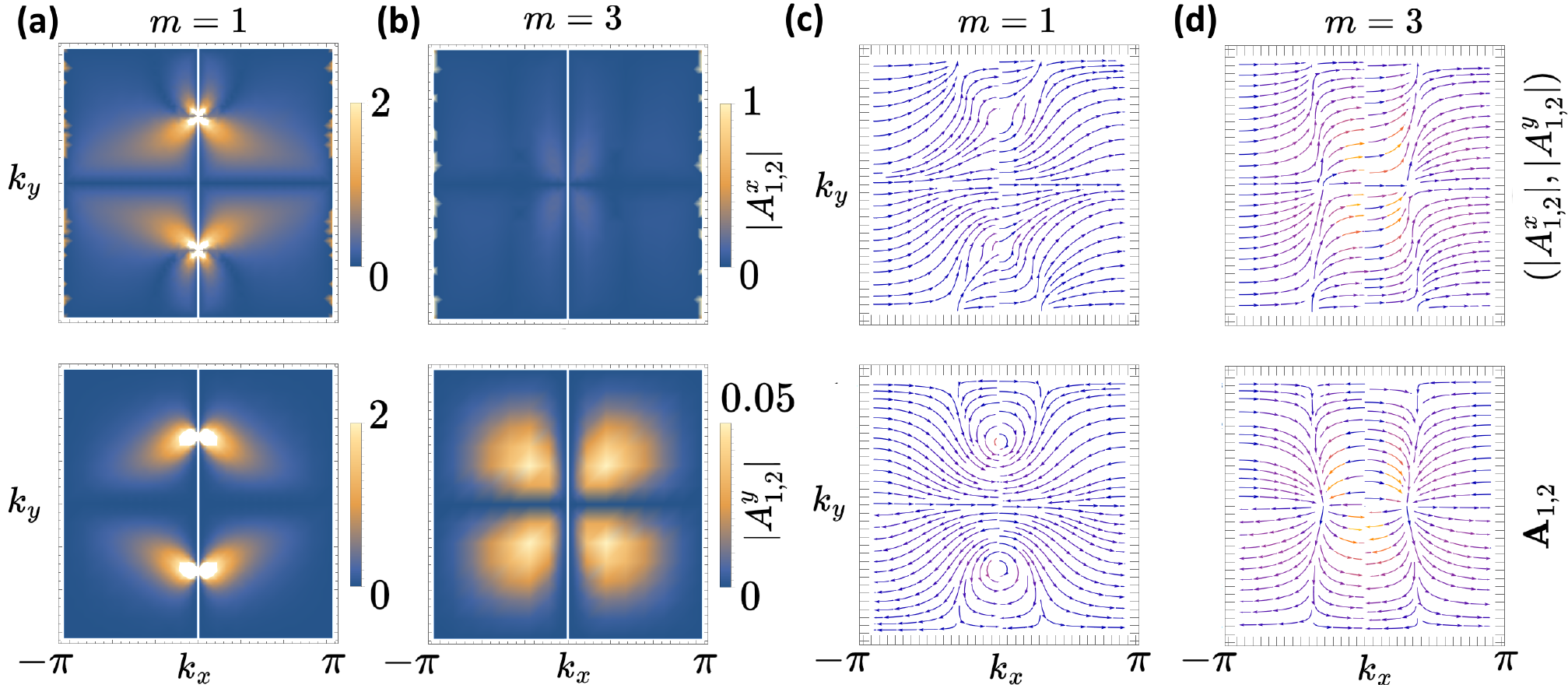}
  \caption{Mapping out the Euler invariant. % from E1 transitions. 
  The absolute values of the components of the non-Abelian connection are reconstructed from the associated quantum metric probed by optical transition rates under linearly polarized light. \textbf{(a)},\textbf{(b)} $|A^x_{12}|$ (upper panel) and $|A^y_{12}|$ (lower panel) are mapped out for the Euler Hamiltonian with $\chi = 2$ on a square lattice, in the topological ($m = 1$) \textbf{(a)} and trivial phase ($m = 3$) \textbf{(b)}. The non-Abelian connection vectors can be reconstructed by fixing the signs of these components in a maximally smooth gauge for topological \textbf{(c)} and  trivial \textbf{(d)} phases.}
\label{fig:reconstruct}
\end{figure}

\sect{Numerical verification in representative models}
Given the general formulation of Euler models using Pl\"ucker embeddings~\cite{Bouhon_2020_geo, bouhon2022multigap}, we can readily corroborate our results with tight-binding models. We consider illustrative three-band and four-band models on square and kagome lattices, which can be synthesized experimentally in trapped-ion simulators and metamaterials~\cite{Unal_2020,Jiang_2021,Jiang_2021}. We present our numerical findings for optical weights in Fig.~\ref{fig:Ws}, demonstrating how the optical bounds from Eq.~\eqref{eq:opticalBound} are satisfied in the topological phases~\cite{SI}. Furthermore, in Fig.~\ref{fig:reconstruct} we show how the Euler invariant can be reconstructed from the optical transitions, as proposed. Finally, we corroborate results from the $\kv \cdot \vec{p}$ model of an Euler node in the lattice-regularized realizations (see SM~\cite{SI}).

\sect{Discussion}
We further comment on the scope and applicability of our results. As derived analytically and verified numerically, our bounds on optical responses offer a route to probe many-band topology in Euler Hamiltonians within the independent-particle picture. We note that these models may be implemented in trapped-ion simulators and metamaterials. Additionally, the bounds on optical weights can be understood microscopically in terms of the application of Fermi's golden rule to interband transitions. In the case of nodal Euler topology, the relation between the optical weight and doping of an Euler semimetal naturally arises, which is sensitive to the value of the patch Euler class. Here, the jerk conductivities, and other third-order effects, provide a valuable insight, especially by means of low-frequency divergences~\cite{PhysRevX.10.041041}. Even when the nodes are not rotationally symmetric, the ratios of photoconductivities obtained within the effective model may be modified. Additionally, on breaking the inversion symmetry assumed in the continuum model, linear injection, as well as circular shift photoconductivities~\cite{SI}, might arise, offering further insights into the physics induced by the non-Abelian connection constrained by the Euler invariant. 

Furthermore, the Euler topology can be removed by breaking $\mathcal{C}_2\mathcal{T}$/$\mathcal{PT}$ symmetry. Upon breaking also $\mathcal{T}$-symmetry individually, a pair of bands with opposite Chern numbers may appear~\cite{bouhon2022multigap}.
In this case, magnetic circular dichroism effects must emerge on doping or frequency tuning~\cite{doi:10.1126/sciadv.1701207,SI}, whereas in Euler phases, such a  dichroism vanishes  by the reality condition.

Finally, the consequences of the derived bound on the gaps below and above the Euler bands can be contrasted with effective models for twisted layered graphene systems~\cite{PhysRevLett.122.106405, Bernevig_2021} assuming flat bands with ${\chi = 1}$, as well as with the single-gap Chern bounds in moir\'e MoTe$_2$~and~WTe$_2$~\cite{onishi2023fundamental}. Contrary to the single-gap condition for Chern insulators~\cite{onishi2023fundamental}, we show that the Euler topology \textit{necessarily} constrains two neighboring gaps through a multigap quantum-geometric bound, Eq.~\eqref{eq:gapbound}. This distinction can be traced back to the fact that the Euler invariants of occupied two-band subspaces are not additive, unlike the Chern numbers of the individual occupied bands, allowing the corresponding bound to be reduced to a single-gap condition. Consistently, our general optical bounds, Eqs.~\eqref{eq:opticalBound},\eqref{eq:FGRbound}, combine two distinct, gapped band occupation configurations ($\text{unocc}$/$\text{occ}$). All bounds thus reflect a multigap nature.

\sect{Conclusions}
We demonstrated how the topological Euler class can be optically probed in electronic systems. In particular, it can be exactly reconstructed from the optical absorption/excitation experiments. Using quantum geometry, we showed that the presence of the two-band subspace hosting an Euler invariant enforces a lower bound on the optical weights and interband contributions to dc conductivities. Furthermore, we derived a fundamental bound constraining a harmonic mean of both gaps neighboring the Euler bands. Finally, we showed that the Euler class of a non-Abelian semimetal can be reconstructed from (quantized) optical conductivities, optical weights, and higher-order photoconductivities, providing feasible avenues for experimental implementations.

\begin{acknowledgements}
    W.J.J.~acknowledges funding from the Rod Smallwood Studentship at Trinity College, Cambridge. A.S.M.~acknowledges funding from EPSRC PhD studentship (Project No. 2606546).  A.B.~has been partly funded by a Marie Sklodowska-Curie fellowship, Grant No. 101025315, and acknowledges financial support from the Swedish Research Council (Vetenskapsradet) (2021-04681). R.-J.S. acknowledges funding from a New Investigator Award, EPSRC Grant No. EP/W00187X/1, a EPSRC ERC underwrite Grant No. EP/X025829/1, and a Royal Society exchange Grant No. IES/R1/221060 as well as Trinity College, Cambridge.  F.N.\"U.~acknowledges funding from the Royal Society via Grant No, URF/R1/241667, the Marie Sk{\l}odowska-Curie programme of the European Commission Grant No.~893915, Simons Investigator Award (Grant No.~511029), Trinity College Cambridge, and thanks the Aspen Center for Physics for their hospitality, where this work was partially funded by a grant from the Sloan Foundation.
    R.-J.S. acknowledges helpful discussions with Prof. E.~J.~Mele and J.~E.~Moore. W.J.J.~thanks Gaurav Chaudhary and Jan Behrends for useful discussions. 
\end{acknowledgements}

\bibliography{references}% Produces the bibliography via BibTeX.
%%%%%%%%%%%%%%%%%%%%%%%%%%%%%%%%%%%%%%%%%%%%%%%%%%%%%%%%%%%%

\newpage

\end{document}

\title{SUPPLEMENTAL MATERIAL \\ Optical manifestations and bounds of topological Euler class}

\author{Wojciech J. Jankowski}
\email{wjj25@cam.ac.uk}
\thanks{}
\affiliation{TCM Group, Cavendish Laboratory, Department of Physics, J J Thomson Avenue, Cambridge CB3 0HE, United Kingdom}

\author{Arthur S. Morris}
\thanks{}
\affiliation{TCM Group, Cavendish Laboratory, Department of Physics, J J Thomson Avenue, Cambridge CB3 0HE, United Kingdom}

\author{Adrien Bouhon}
\thanks{}
\affiliation{TCM Group, Cavendish Laboratory, Department of Physics, J J Thomson Avenue, Cambridge CB3 0HE, United Kingdom}
\affiliation{Nordita, Stockholm University and KTH Royal Institute of Technology, Hannes Alfv{\'e}ns v{\"a}g 12, SE-106 91 Stockholm, Sweden}

 \author{F. Nur \"Unal}
\thanks{}
\affiliation{TCM Group, Cavendish Laboratory, Department of Physics, J J Thomson Avenue, Cambridge CB3 0HE, United Kingdom}
\affiliation{School of Physics and Astronomy, University of Birmingham, Edgbaston, Birmingham B15 2TT, United Kingdom}%\looseness=-1}

\author{Robert-Jan Slager}
\email{rjs269@cam.ac.uk}
\affiliation{TCM Group, Cavendish Laboratory, Department of Physics, J J Thomson Avenue, Cambridge CB3 0HE, United Kingdom}
\affiliation{Department of Physics and Astronomy, University of Manchester, Oxford Road, Manchester M13 9PL, United Kingdom}

\date{\today}

\maketitle

\section{A brief introduction to quantum geometry}\label{app:appA}

\subsection{Bounds on multiband invariants}
In the following we briefly review elements of the non-Abelian quantum geometry of multiband systems. To do so we will make use of the recently established Pl\"ucker formalism, which allows for a concrete analysis of quantum geometry beyond the projective single-band description \cite{bouhon2023quantum}. We further utilize this tool to describe the geometric quantities present in optical response in terms of Riemannian metrics; in a similar spirit to Refs.~\cite{PhysRevX.10.041041, Ahn_2021}. 

We define the \textbf{non-Abelian quantum geometric tensor} componentwise as \cite{bouhon2023quantum}
\begin{subequations}\label{eq:NA_qu_GTs}
\beq{eq:NA_qu_GT}
    (\tilde{Q}_{ab})_{nm} = \bra{\partial_a u_{n,\kv}} \hat{Q}_{l,\kv} \ket{\partial_b u_{m,\kv}} = \bra{\partial_a u_{n,\kv}} 1 - \hat{P}_{l,\kv} \ket{\partial_b u_{m,\kv}},
\eeq
where $\hat{P}_{l,\kv} = \sum_{\ell\in l} \ket{u_{\ell,\kv}}\bra{u_{\ell,\kv}}$ is the projector onto the manifold spanned by a subset $l\subset \{1, \ldots, N\}$ of selected bands of interest \cite{PhysRevLett.124.167002, bouhon2023quantum}, which need not contain the entire manifold of occupied states. Here, as in the main text, the indices $a, b = k_x, k_y$ label momenta, while $n, m \in l$ run over the chosen bands. The \textbf{non-Abelian quantum metric} corresponds to the real part of this tensor,
\beq{eq:NA_qu_met}
    (\tilde{g}_{ab})_{nm} = \Re (\tilde{Q}_{ab})_{nm} = \Re \bra{\partial_a u_{n,\kv}} \hat{Q}_{l,\kv} \ket{\partial_b u_{m,\kv}},
\eeq
while the \textbf{non-Abelian Berry curvature} is ($-2$ times) the imaginary part, that is 
\beq{eq:NA_qu_C}
\mathcal{F}^{nm}_{ab} = -2\Im (\tilde{Q}_{ab})_{nm} = -2\Im \bra{\partial_a u_{n,\kv}} \hat{Q}_{l,\kv} \ket{\partial_b u_{m,\kv}}. 
\eeq
\end{subequations}
Eqs. \eqref{eq:NA_qu_GTs} may be written in the combined form,
\beq{}
    (\tilde{Q}_{ab})_{nm} = (\tilde{g}_{ab})_{nm} -\frac{\ii}{2} \mathcal{F}^{nm}_{ab}
\eep
Importantly, $\tilde{Q}_{ab}$ is \textit{positive semidefinite}, as is the non-Abelian quantum metric  $\tilde{g}_{ab}$~\cite{PhysRevLett.124.167002, bouhon2023quantum}. 

For any choice of bands $l$, one may define an \textbf{Abelian quantum metric} $g^l_{ab}$ by taking the trace of $\tilde{g}_{ab}$ over these bands:
\begin{equation}\label{eq:gll}
    g^l_{ab} \equiv \Tr_l \tilde{g}_{ab} = \sum_{n \in l} (\tilde{g}_{ab})_{nn}.
\end{equation}
In particular, when $l$ corresponds to all occupied bands $l_\text{occ}$, one obtains the Abelian quantum metric $g_{ab} \equiv g^{l_\text{occ}}_{ab}$, as defined in literature~\cite{Ahn_2021}.
For an arbitrary subset $l$ of bands it can be seen that $g^l_{ab}$ is also a Fubini-Study metric, and may be used to assign a `distance' between the infinitesimally separated points $\kv$ and $\kv + \dd \kv$ as \cite{T_rm__2018, PhysRevLett.124.167002, bouhon2023quantum}
\beq{}
    D^2(\kv, \kv + \dd \kv) = \frac{1}{2} \Tr\left[ \left(\hat{P}_{l,\kv + \dd \kv} - \hat{P}_{l,\kv} \right)^2 \right]= g^l_{ab}(\kv)\, dk_a\, dk_b.
\eeq
We now turn to the relation of the non-Abelian quantum geometric tensor to multiband topological invariants. To investigate the Euler topology of a subspace spanned by two bands $\ket{1}\equiv\ket{u_{1, \kv}}$ and $\ket{2}\equiv\ket{u_{2, \kv}}$ we set $l = \{1, 2\}$ to consist of these bands only.
In this case it can be seen that the Euler curvature, $\text{Eu} = -\mathcal{F}^{12}_{xy}$, appears directly within the quantum geometric tensor. The positive semidefiniteness of the geometric tensor in fact leads to the inequality condition on the Euler class that is used in the main text; we will now derive this equation. Using the shorthand notation $\ket{\ell,a} \equiv \ket{\partial_a u_{\ell,\kv}}$, where $\ell\in l$, we may write
\begin{subequations}
\beq{}
    (\tilde{g}_{ab})_{nm} = \bra{n,a} \hat{Q}_{l,\kv} \ket{m, b}
\eeq
and
\beq{}
    \text{Eu} = \bra{1,y} \hat{Q}_{l,\kv} \ket{2,x} - \bra{1,x} \hat{Q}_{l,\kv} \ket{2,y}.
\eeq
\end{subequations}
Applying the quantum geometric tensor to a judiciously chosen set of states $\ket{\phi_1^{\pm}} = \frac{1}{\sqrt{2}}(\ket{1,y} \pm \ket{2,x})$ and $\ket{\phi_2^{\pm}} = \frac{1}{\sqrt{2}}(\ket{1,x} \pm \ket{2,y})$
then leads to a set of relations
\beq{}
    \bra{\phi^{\pm}_i}\hat{Q}_{l,\kv} \ket{\phi^{\pm}_i} \geq 0, \qquad i = 1, 2
\eeq
which, by using the relation $\bra{n,i}\ket{m,j} = \bra{m,j}\ket{n,i}$ from the reality condition, may be rearranged to yield
\beq{}
    \Tr \textbf{g}^{\chi} \equiv (\tilde{g}_{xx})_{11} + (\tilde{g}_{yy})_{11} + (\tilde{g}_{xx})_{22} +(\tilde{g}_{yy})_{22} \geq \pm 2 \text{Eu},
\eeq
or, more compactly, $\Tr \textbf{g}^{\chi} \geq 2|\text{Eu}|$. Hence, upon performing the integration over the BZ we arrive at the required inequality
\beq{eq:ramn}
    \frac{1}{4\pi} \int_{\text{BZ}} \dd^2 \kv~ \Tr \textbf{g}^{\chi} \geq \left| \frac{1}{2\pi} \int_{\text{BZ}} \dd^2 \kv~ \mathrm{Eu} \right| = |\chi|,
\eeq
which is central to the main text.

\subsection{Riemannian geometry of Bloch states}
In the context of optical responses relevant to this work, following Ref.~\cite{Ahn_2021}, we identify the transition dipole matrix element $\rv_{mn}(\kv) = \bra{\psi_{m\kv}} \hat{\rv} \ket{\psi_{n\kv}}$ as a tangent vector on the manifold of quantum states, with components
\beq{}
    \rv^a_{mn}(\kv) = \ii\Big(\delta_{mn}\partial_a - A^{a}_{n,m}(\kv)\Big)
\eec
where $A^{a}_{n,m}(\kv) \equiv \bra{u_{n,\kv}}  \ket{\partial_{a} u_{m,\kv}}$ is the non-Abelian Berry connection, and  $\partial_a \equiv \frac{\partial}{\partial_{k_a}}$ is a tangent vector component induced by the local coordinates $\{ k_a \}$. The manifold upon which these objects are defined depends upon the physical context. For example, if no degeneracies are present then the manifold of states for an $N$-band system is a complex flag manifold $\widetilde{\mathsf{Fl}}_{1,1,\ldots, 1}(\mathbb{C}) = U(N)/U(1)^N$, where the quotient is due to the gauge freedom in each band. In the case of Euler phases, or more general multigap topology under the symmetry-enforced reality conditions central to this work, the manifold of interest extends to: $\widetilde{\mathsf{Fl}}_{p_1, \ldots, p_k}(\mathbb{R}) = O(N)/(O(p_1) \cross \ldots \cross O(p_k))$, where the number of bands in isolated subspaces $k$ sums to $N = p_1 + \ldots + p_k$.

Following Ref.~\cite{Ahn_2021}, we define a Hilbert-Schmidt inner product of matrices as,
\beq{}
    \langle A, B \rangle = \Tr  \left[ A^\dagger B \right] = \sum_{a,b} A^*_{ab} B_{ab}
\eec
which allows the Abelian quantum geometric tensor to be represented in terms of the local interband transition vectors $\hat{e}^{mn}_a(\kv) = \rv^a_{mn}(\kv) \ket{u_{m,\kv}} \bra{u_{n,\kv}}$ as
\beq{}
    Q^{mn}_{ab} \equiv \langle \hat{e}^{mn}_a, \hat{e}^{mn}_b \rangle = r^a_{nm}r^b_{mn}.
\eeq{}
The advantage of this representation is that it allows to easily introduce a Hermitian connection as
\beq{}
    C^{mn}_{acb} \equiv \langle \hat{e}^{mn}_a, \nabla_c \hat{e}^{mn}_b \rangle = r^a_{nm}r^{b;c}_{mn} = \sum_d Q^{mn}_{ad} (C^{mn})^d_{cb},
\eeq{}
where the covariant derivative acts on the transition vectors as $\nabla_c \hat{e}^{mn}_a = \sum_b (C^{mn})^b_{ca} \hat{e}^{mn}_b$. The covariant derivative $r^{b;c}_{nm}$ of the transition dipole is defined in terms of the Berry connection as
\\
\beq{cov}
\begin{split}
    r^{c;a}_{nm} (\kv) = \nabla_{a} r^{c}_{nm} (\kv) = \partial_{a} r^{c}_{nm} (\kv) - \ii(A^{a}_{n,n} (\kv) - A^{a}_{m,m} (\kv)) r^{c}_{nm} (\kv).
\end{split}
\eeq
Correspondingly, the Christoffel symbols for the metric connection are given by
\beq{}
    \Gamma^{mn}_{abc} = \frac{1}{2} (\partial_c g^{mn}_{ba} + \partial_b g^{mn}_{ca} - \partial_a g^{mn}_{bc}) = \Re \left[ r^a_{nm}r^{b;c}_{mn} \right] ,
\eeq
while the torsion tensor is defined as $T^{mn}_{abc} = C^{mn}_{abc} - C^{mn}_{acb}$, and the symplectic connection can be defined as ${\tilde{\Gamma}_{acb} = \ii(C_{acb}-C^*_{acb})}$. Finally, we define the Hermitian curvature tensor as 
\beq{}
\begin{split}
    K^{mn}_{abcd} \equiv \left\langle \hat{e}^{mn}_a, (\nabla_c \nabla_d - \nabla_d \nabla_c) \hat{e}^{mn}_b \right\rangle,
\end{split}
\eeq
and moreover we define the Riemannian curvature tensor as the real part of this object, that is,  $R^{mn}_{abcd} = \Re \{ K^{mn}
_{abcd} \}$. This completes the glossary of quantum Riemannian objects relevant to this work.

\section{Optical responses and quantum geometry}\label{app:appB}

The geometric quantities introduced in the previous section are physically related to the experimentally accessible  \textbf{first-order optical conductivity}~\cite{Ahn_2021}
\begin{equation}\label{eq:AC_OptCond}
    \sigma^{ab} = \frac{\pi \om e^2}{\hbar} \sum_{m,n} \int \frac{\dd^d k}{(2\pi)^d} \delta(\om - \om_{mn}) f_{nm} Q^{mn}_{ab},
\end{equation}
where, ${\om = \om_{mn} = \frac{1}{\hbar}(E_{m,\kv}-E_{n,\kv})}$ is the resonant frequency for transition, and $f_{mn} = f_{n\kv}- f_{m\kv}$ is the difference in Fermi-Dirac occupation factors. More directly, the interband contribution to the imaginary part of the dielectric tensor, which encodes optical absorption, explicitly contains the quantum metric:
\beq{}
    \epsilon^{\prime \prime}_{ab}(\omega) = \frac{\pi e^2}{\hbar} \sum_{m,n} \int \frac{\dd^d k}{(2\pi)^d} \delta(\om - \om_{mn}) f_{nm} g^{mn}_{ab}
\eep
Higher-order optical responses contain a plethora of quantum geometric quantities \cite{Ahn_2021}: the \textbf{second-order shift photoconductivity} is
\begin{subequations}
\beq{}
    \sigma^{cab}_{\text{shift}} = -\frac{\pi e^3}{2\hbar^2} \sum_{m,n} \int \frac{\dd^d k}{(2\pi)^d} \delta(\om - \om_{mn}) f_{nm} \ii\left(C^{mn}_{acb}-(C^{mn}_{bca})^{*}\right),
\eeq
and the \textbf{second-order injection photoconductivity} is
\beq{}
    \sigma^{cab}_{\text{inj}} = -\frac{\pi e^3 \tau}{\hbar^2} \sum_{m,n} \int \frac{\dd^d k}{(2\pi)^d} \delta(\om - \om_{mn}) f_{nm} Q^{mn}_{ab} \partial_c \om_{mn},
\eeq
where $\tau$ is the relaxation time for the photo-excited particle to decay; furthermore, the \textbf{third-order jerk photoconductivity} is given by 
\beq{}
    \sigma^{cdab}_{\text{inj}} = -\frac{2\pi e^4 \tau}{\hbar^3} \sum_{m,n} \int \frac{\dd^d k}{(2\pi)^d} \delta(\om - \om_{mn}) f_{nm} Q^{mn}_{ab} \partial_c \partial_d \om_{mn}.
\eeq
\end{subequations}
One may also define a \textbf{third-order injection photoconductivity}, which depends on $K^{mn}_{cbad}$ \cite{Ahn_2021}, as well as a \textbf{third-order shift photoconductivity} \cite{PhysRevB.100.064301}. Note that the second-order shift conductivity does \textit{not} require any dispersion in the band structure, contrary to the injection and jerk conductivities. These correspondingly depend on the group velocity matrix elements (${v^c_{mm} - v^c_{nn} = \partial_c \om_{mn}}$) and effective masses (${(m^{-1}_*)_{cd,m}- (m^{-1}_*)_{cd,n} = \frac{1}{\hbar}\partial_c \partial_d \om_{mn}}$) in bands $m,n$ between which the photo-excitation occurs. 

The optical conductivity can be decomposed as
\beq{}
\sigma^{ab}(\om)   = \Re \sigma^{ab}(\om) + \ii \Im~\sigma^{ab}(\om),
\eeq
and we define generalized \textbf{optical weights} \cite{onishi2023fundamental} as
\beq{}
    W^\kappa_{ab}(\om_{\text{max}}) = \int_{0}^{\om_{\text{max}}} \dd \om ~\frac{\Re \sigma^{ab}(\om)}{\om^\kappa}.
\eeq
In particular, for $\kappa = 1$,
\beq{opt_metric}
\begin{split}
    W^1_{ab}(\om_{\text{max}}) = \int_{0}^{\om_{\text{max}}} \dd \om ~\frac{\Re  \, \sigma^{ab}(\om)}{\om} = \frac{\pi e^2}{\hbar} \sum_{m,n} \int \frac{\dd^d k}{(2\pi)^d} \theta(\om_{\text{max}} - \om_{mn}) f_{nm} g^{mn}_{ab},
\end{split}
\eeq
where $\theta(\om_{\text{max}} - \om_{mn})$ is the Heaviside step function, which acts to select contributions up to a maximum frequency $\om_{\text{max}}$. Using the Kramers-Kronig relations
\begin{subequations}
\begin{align}
    \frac{2}{\pi} P \int_{0}^{\infty} \dd \om ~\frac{\Re\sigma^{ab}(\om)}{\om-\om'} ={}& \Im\sigma^{ab}(\om'), \label{eq:KK1}\\ \label{eq:KK2}
    -\frac{2}{\pi} P \int_{0}^{\infty} \dd \om ~\frac{\Im \sigma^{ab}(\om)}{\om-\om'} ={}& \Re\sigma^{ab}(\om'),
\end{align}
\end{subequations}
 along with the identity $\sigma^{ab}(\om) = \sigma^{ab}(-\om)^*$, and setting $\om' = 0$, we observe that  
\beq{eq:interbandDC}
    \Im\sigma^{ab}(0) 
    = \frac{2 e^2}{\hbar} \sum_{m,n} \int \frac{\dd^d k}{(2\pi)^d} f_{nm} g^{mn}_{ab}= \sigma_{\text{inter}}^{ab}(0),
\eeq
Where we have identified $\sigma_{\text{inter}}^{ab}(0)$, the quantum interband contribution to the dc longitudal conductivity \cite{PhysRevB.102.165151, PhysRevB.105.085154}. 
Similarly, the real part of the dc conductivity is
\beq{}
    \Re\sigma^{ab}(0) 
    = -\frac{2 e^2}{\hbar} \sum_{m,n} \int \frac{\dd^d k}{(2\pi)^d} f_{nm} \left( -\frac{F^{mn}_{ab}}{2} \right).
\eeq
This identity contains as a special case the famous TKNN formula \cite{PhysRevLett.49.405} 
\beq{}
\begin{split}
    \sigma_{xy} = \Re \sigma^{xy}(0) 
    = \frac{e^2}{\hbar} \sum_{m,n} \int \frac{\dd^2 k}{(2\pi)^2} f_{nm} F^{mn}_{xy} = \frac{e^2}{2\pi\hbar} \sum_{n} \frac{1}{2\pi} \int \dd^2 k~ F^{nn}_{xy} = \frac{e^2}{h} \sum_{n} C_n.
\end{split}
\eeq
which describes the quantization of the (anomalous) Hall conductivity, $\sigma_{xy}$, in two-dimensions. We therefore see that while the second Kramers-Kronig relation Eq.~\ref{eq:KK2} provides information about the Chern topology, it is the other Kramers-Kronig relation Eq.~\ref{eq:KK1} that is central to indicating the presence of Euler topology, as explained in the main text. 

Similarly to the first-order optical conductivity, both the injection and shift conductivities can be decomposed in terms of $\textbf{linear}$ ($L$) and $\textbf{circular}$ ($C$) terms (probed by linearly/circularly-polarized light, respectively) as
\beq{}
    \sigma^{cab}_{\text{inj}} = \sigma^{cab}_{\text{inj},L} + \ii\sigma^{cab}_{\text{inj},C}.
\eeq
For instance, the bulk linear shift photoconductivity for a two-dimensional system is given by \cite{Iba_ez_Azpiroz_2018}:

\beq{}
\begin{split}
    \sigma^{cab}_{\text{shift},L}(\om) = \frac{\pi e^3}{2\hbar^2} \int \frac{\dd^2k}{(2\pi)^2} \sum_{nm} (f_{n\kv}-f_{m\kv}) \Im\big[r^{b}_{nm} (\kv) r^{c;a}_{mn} (\kv) + r^{b}_{nm} (\kv) r^{ca}_{mn} (\kv)\big] \delta(\om_{mn} - \om).
\end{split}
\eeq
It is immediately clear that, under the reality condition required for nontrivial Euler topology, the integral vanishes, and $\sigma^{cab}_{\text{shift},L}(\om) = 0$. Similarly, the vanishing of the Abelian Berry curvature, which follows from the vanishing of the Abelian connection in Euler materials, necessarily implies $\sigma^{cab}_{\text{inj},C}(\om) = 0$. 

The photoconductivities defined above describe the dc responses induced by an applied electric field; such responses are of general interest in the context of photovoltaic applications \cite{Cook_2017, alexandradinata2022topological}. We briefly mention a few cases of particular importance: firstly
\begin{subequations}
\beq{}
    j^c_{\text{shift/inj}}(0) = 2 \sum_{a,b} \sigma^{cab}_{\text{shift/inj}}(\om)\mathcal{E}^a(\om)\mathcal{E}^b(-\om),
\eeq
for second order dc currents; at third-order we have
\begin{align}
    j^d_{\text{shift/inj}}(0) ={}& 6 \sum_{a,b,c} \sigma^{dabc}_{\text{shift/inj}}(\om)\mathcal{E}^a(\om)\mathcal{E}^b(-\om)\mathcal{E}^c(0),\\
    j^d_{\text{jerk}}(0) ={}& 6 \sum_{a,b,c} \sigma^{dcab}_{\text{jerk}}(\om)\mathcal{E}^a(\om)\mathcal{E}^b(-\om)\mathcal{E}^c(0),
\end{align}
\end{subequations}
where $E^d(0)$ is an additional dc electric field component. Importantly, different photoconductivities can be probed with both linearly and circularly-polarized light, inducing corresponding dynamical responses.

Finally, as shown in  Ref.~\cite{Ahn_2021}, we note that there also exists a manifestation of the last geometric quantity introduced in the previous section, namely the Riemann curvature tensor $R^{mn}_{cbad}$,
\beq{}
    \chi_{m,n} = C_m - C_n = \frac{1}{2\pi} \oint_{\text{BZ}} \dd^2k  \frac{\text{sgn}(-\Im Q^{mn}_{12}) R^{mn}_{1212}}{\sqrt{\text{det}(g^{mn})}}
\eec
in a topological optical photovoltaic Hall response. This response is captured by so-called `Euler numbers' $\chi_{m,n}$ defined as differences of Chern numbers of bands $m$ and $n$, correspondingly $C_m$ and $C_n$. Importantly, as can be checked by inspection, these Euler numbers are distinct from the Euler class $\chi$ of two neighboring bands with band crossings, which can be viewed as the Chern number of the  complexification of two bands: $\chi (\ket{u_{n}}, \ket{u_{n+1}})  = C\big[ \frac{1}{\sqrt{2}} (\ket{u_{n}} + \ii \ket{u_{n+1}})\big] $, rather than as the difference of their Chern numbers.

%By instead expressing the Chern numbers of each band in terms of the Euler class of the complexified bands, we can obtain a relationship between these two quantities: 
%\begin{align}
%\begin{split}
%    \chi_{n,n+1} ={}& C_n - C_{n+1}\\
%    ={}& \chi\left(\frac{\ket{u_{n}}+\ket{u_{n+1}}}{\sqrt{2}}, \frac{\ket{u_{n}}-\ket{u_{n+1}}}{i\sqrt{2}}\right) - \chi\left(\frac{\ket{u_{n}}+\ket{u_{n+1}}}{\sqrt{2}}, \frac{\ket{u_{n+1}}-\ket{u_{n}}}{i\sqrt{2}}\right).
%\end{split}
%\end{align}

\section{Minimal lattice-regularized Euler models}\label{app:appC}

The minimal patch Euler class $\chi = 1$ can be realized in a kagome lattice model, where there are three orbitals per unit cell, by including nearest neighbour (NN) $(t)$, next-nearest neighbour (NNN) $(t')$, and third-nearest neighbour (N3) hoppings $(t'')$~\cite{Jiang_2021}. The corresponding tight-binding Hamiltonian can be written as:
\begin{equation}
\label{eq:kagomeH}
    H = \sum_i \varepsilon_i c^\dagger_i c_i + t \sum_{\langle ij \rangle}(c^\dagger_i c_j + \textnormal{h.c.}) %\\
    + t' \sum_{\langle\langle ij \rangle\rangle}(c^\dagger_i c_j + \textnormal{h.c.}) + t''\sum_{\langle\langle\langle ij \rangle\rangle\rangle}(c^\dagger_i c_j + \textnormal{h.c.}).
\end{equation}
A variety of phases can be realized within this model. Firstly, setting the onsite energies of all orbitals to $\varepsilon_i = 0$ and choosing vanishing N3 hoppings $t'' = 0$ yields a quadratic Euler node on choosing, e.g., $t=1$, $t'=-0.2$. A gapped model with $\chi = 1$ can be obtained on further setting $t=1$, $t'=0$, Fourier transforming the Hamiltonian, and adding a diagonal term $\text{diag}(-3,-3,0)$.

%For higher Euler invariants, we introduce the following, general three-band Hamiltonians of arbitrary Euler class
%\\
%\beq{Model}
%    H(\kv) = \exp(\ii \vb{L}\cdot\vb{}{n}(\kv))
%\eec
%\\
%where $\vb{L}=(L_x, L_y, L_z)$ is the vector of generators of the Lie algebra $\mathfrak{so}(3)$, 
Additionally, we also investigate the following orientable three-band models with arbitrarily high even Euler class $\chi = 2 p_x p_y$, set by $p_x, p_y \in \mathbb{Z}$ as:
\beq{Even-Euler}
    H^{[\chi]} = 2 \ket{u_3} \bra{u_3} - \mathbb{I}_3,
\eeq{}
which was realized experimentally in an optical lattice and trapped-ion setups \cite{Unal_2020,Zhao_2022}; hence we focus on this class of models in the main text.
Here,
$\textbf{n}_3 (\kv) \equiv \textbf{u}_1 (\kv) \times \textbf{u}_2 (\kv)$ is the Bloch eigenvector corresponding to the third band. To induce winding underlying the nontrivial Euler class, we choose 
\\
\beq{}
\textbf{n}_3 (\kv) = \frac{1}{\mathcal{N}}
\begin{pmatrix}
\sin p_x k_x \\ 
\sin p_y k_y \\
m - \cos p_x k_x - \cos p_y k_y \\
\end{pmatrix}
\eeq
\\
where $m$ denotes a topological mass term and $\mathcal{N}$ is a normalization factor. The Euler invariant in any three-band model can be expressed as
\beq{}
    \chi = \frac{1}{4\pi} \int_{\text{BZ}} \dd^2k~\textbf{n}_3 \cdot(\partial_{k_x} \vb{n}_3 \times \partial_{k_y} \textbf{n}_3)
\eec
\\
which equivalently can be written in terms of the Euler 
{2-form} ${\text{Eu} = \left[ \bra{\partial_{k_x}u_1}\ket{\partial_{k_y}u_2} - \bra{\partial_{k_y}u_1}\ket{\partial_{k_x}u_2} \right]\, \dd k_x \wedge \dd k_y}$ as ${\chi = \frac{1}{2\pi} \int_{\text{BZ}}} \text{Eu}.$ 

Further to this, we study four-band models with \textit{double} Euler class, denoted $(\chi_1, \chi_2)$. Specifically, we utilize a Hamiltonian introduced in Ref.~\cite{bouhon2022multigap}, 
\begin{align}
    H^{[\chi_1, \chi_2]}(\textbf{k}) =\sine{k_1}\Gamma_{01} + \sine{k_2}\Gamma_{03}
   - [m - t_1(\cosine{k_1} + \cosine{k_2}) - t_2\cosine{(k_1+k_2)}]\Gamma_{22} + \delta\Gamma_{13},
\end{align}
parameterized by the variables $(m, t_1, t_2, \delta)$, where the $\Gamma_{ij} = \sigma_{i} \otimes \sigma_{j}$ are $4 \cross 4$ matrices. Representative models for $(\chi_1, \chi_2) = (1, 1)$ and $(\chi_1, \chi_2) = (2, 2)$ can be obtained by setting the parameters $(m, t_1, t_2, \delta)$ to $(1,-3/2, 0, 1/2)$ and $(1/2,-1/2, -3/2, 1/2)$ respectively.

\section{Derivation of the bound on the combined optical weights in Euler materials}\label{app:appD}
In this Section we derive the bound on the combined optical weights at zero temperature due to the nontrivial Euler class. We consider a system with a total of $N$ bands (in particular a limit of infinitely, but countably many bands $N \rightarrow \infty$ can be taken), where the bands $M+1$ and $M+2$ (with
$M+2\leq N$) together form a nontrivial Euler subspace. With reference to Eq. \ref{eq:gll}, we define three distinct quantum metrics in this space by specifying the bands $l$ with which they are associated. Firstly, we define the metric of the Euler bands only as $g^{\chi}\equiv g^{l_{\chi}}$, where $l_{\chi}=\{ M+1, M+2\}$. Secondly, we suppose that the chemical potential of the system sits between the bands $M+2$ and $M+3$, so that the Euler bands are occupied, and define $g_{\text{occ}}\equiv g^{l_{\text{occ}}}$, with $l_{\text{occ}}=\{1, \ldots, M, M+1, M+2\}$. Finally, we define a metric for the case that the chemical potential sits between bands $M$ and $M+1$, so that the Euler subspace is unoccupied, and $g_{\text{unocc}}\equiv g^{l_{\text{unocc}}}$ with $l_{\text{unocc}}=\{1, \ldots, M\}$. For later convenience we also let $l_0=\{1, \ldots, N\}$ and define $\bar{l}_{\text{occ}}=l_0\backslash l_{\text{occ}} = \{M+3, \ldots, N\}$, $\bar{l}_{\text{unocc}}=l_0\backslash l_{\text{unocc}} = \{M+1, \ldots, N\}$, and $\bar{l}_\chi =l_0\backslash l_{\chi}= \{1, \ldots, M, M+3, \ldots, N\}$. Note that $\bar{l}_\chi = l_{\text{unocc}}\cup \bar{l}_{\text{occ}}$, $l_{\text{occ}} = l_\chi\cup l_{\text{unocc}}$, and $\bar{l}_{\text{unocc}} = l_\chi\cup \bar{l}_{\text{occ}}$. Using the definition Eq. \ref{eq:NA_qu_GTs}, we may then write each of these metrics in terms of the Berry connection as follows:
\begin{subequations}
    \begin{align}
        [g_{\chi}]_{aa} ={}& \sum_{n\in l_{\chi}}\sum_{m\in \bar{l}_{\chi}}|A^a_{nm}|^2 =  \sum_{n\in l_\chi} \sum_{m\in \bar{l}_{\text{occ}}} |A^a_{nm}|^2 + \sum_{n\in l_{\chi}} \sum_{m\in l_{\text{unocc}}} |A^a_{nm}|^2 \equiv X_1+X_2,\\
        [g_{\text{occ}}]_{aa} ={}& \sum_{n\in l_{\text{occ}}}\sum_{m\in \bar{l}_{\text{occ}}}|A^a_{nm}|^2 =  \sum_{n\in l_\chi} \sum_{m\in \bar{l}_{\text{occ}}} |A^a_{nm}|^2 + \sum_{n\in l_{\text{unocc}}} \sum_{m\in \bar{l}_{\text{occ}}} |A^a_{nm}|^2\equiv O_1+O_2,\\
        [g_{\text{unocc}}]_{aa} ={}& \sum_{n\in l_{\text{unocc}}}\sum_{m\in \bar{l}_{\text{unocc}}}|A^a_{nm}|^2 =  \sum_{n\in l_{\text{unocc}}} \sum_{m\in l_{\chi}} |A^a_{nm}|^2 + \sum_{n\in l_{\text{unocc}}} \sum_{m\in \bar{l}_{\text{occ}}} |A^a_{nm}|^2\equiv U_1+U_2.
    \end{align}
\end{subequations}
It can be immediately seen that $X_1=O_1$ and $O_2=U_2$, and, since $|A^a_{nm}|^2 = |A^a_{mn}|^2$, we have also $X_2=U_1$. Since all terms in these equations are non-negative sums of squares, it is also apparent that $X_1, X_2, O_1, O_2, U_1, U_2\geq 0$. It thus follows that 
\begin{equation}
    [g_{\text{occ}}]_{aa}+[g_{\text{unocc}}]_{aa} = O_1+O_2+U_1+U_2 = X_1+X_2+ O_2 + U_2\geq X_1+X_2 = [g_{\chi}]_{aa}.
\end{equation}
By summing over the coordinate index $a=x, y$, integrating over the BZ, and using Eqs.~\eqref{eq:ramn} and ~\eqref{opt_metric}, we arrive at
\beq{}
    W^1_{xx,(\text{occ})}(\infty) + W^1_{yy,(\text{occ})}(\infty) + W^1_{xx,(\text{unocc})}(\infty) + W^1_{yy,(\text{unocc})}(\infty) \geq \frac{ e^2}{\hbar} |\chi|,
\eeq
as required. Physically, we recognize that all terms introduced above correspond to E1 transition matrix elements, and $O_2 = U_2$ represents the background of all transitions which are not to/from the two Euler bands. While this background consists of the matrix elements for all transitions from below the Euler bands to above them; $O_1 = X_1$ represents all transitions from within the Euler bands to all bands above, while $U_1 = X_2$ is the sum over transitions from all available bands below the Euler bands to the two specified Euler bands.

\begin{figure}
\centering
%  \captionsetup[figure]{justification=justified, singlelinecheck=off} 
  \includegraphics[width=0.8\columnwidth]{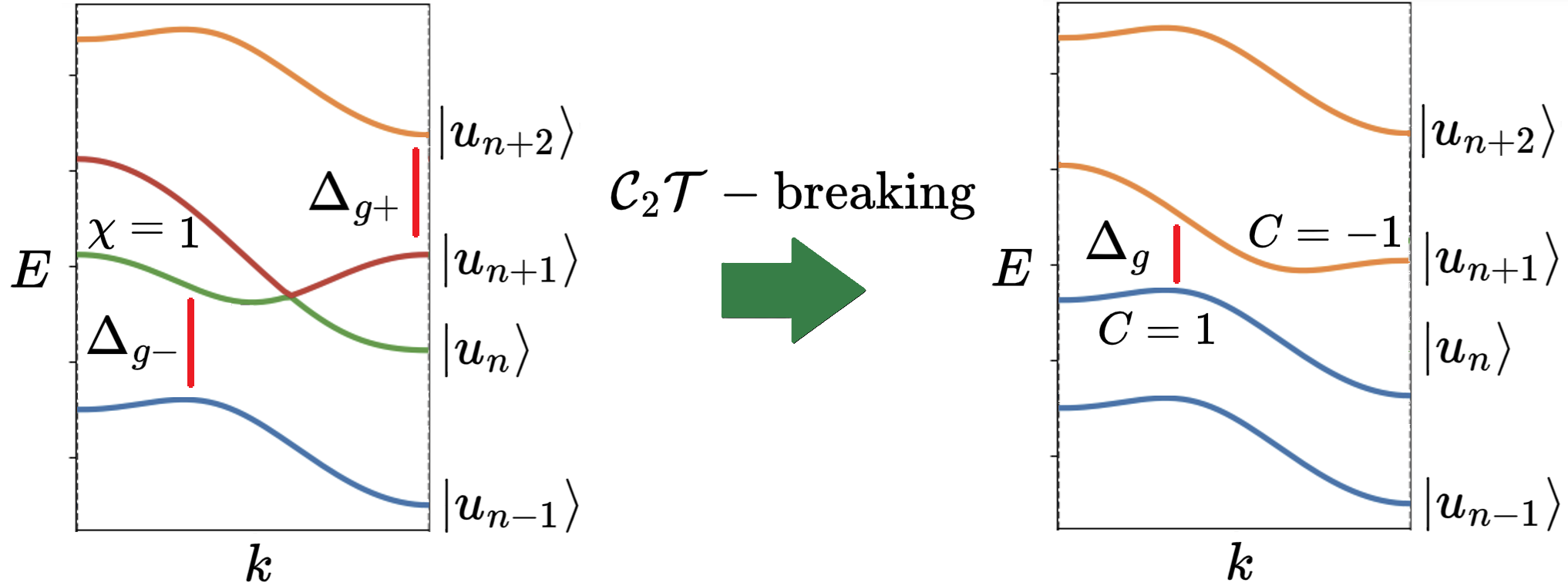}
  \caption{Comparison of topological bounds due to single-gap and multigap conditions. On breaking the $\mathcal{C}_2\mathcal{T}$ symmetry, the pair of bands with an Euler class $\chi$ can be transformed into a pair of Chern bands with Chern numbers $C = \pm \chi$. As a result, the gaps below the valence Chern and above the conduction Chern bands can be closed now [with trivial bands below and above] without trivializing the Chern topology of the ground state. In the case of Euler topology, the gaps below and above the Euler band need to be present in order to keep the nontrivial Euler invariant of a pair of bands over the entire Brillouin zone. Here, closing the gaps with \textit{trivial} bands above and below the two-band Euler subspace, trivializes the fragile Euler invariant. The topologically bounded gaps are marked with red vertical lines. In this work, we retrieve a bound on the gaps above an below the Euler bands $\Delta_{g+}$ and $\Delta_{g-}$, which we contrast with the bound on the gap $\Delta_g$ between the descendent Chern bands that was derived in Ref.~\cite{onishi2023fundamental}.}
\label{fig:S1}
\end{figure}

For completeness -- we now further comment on the differences between the optical bounds due to the Euler band topology and Chern topology~\cite{onishi2023fundamental}. We illustrate the distinction with Fig.~\ref{fig:S1}. Here, we further clarify and discuss these differences with a number of arguments that revolve around the context of topological phase transitions between Euler and Chern bands on breaking $\mathcal{C}_2\mathcal{T}$ symmetry, as mentioned in the Discussion section of the main text.

{\it (i)} Breaking the $\mathcal{C}_2\mathcal{T}$ symmetry and opening the gap between the Euler bands to obtain two opposite Chern bands~\cite{bouhon2022multigap} introduces a single gap bound~\cite{onishi2023fundamental} on the excitations across the emerging gap, not on the gap below and above as in the case of the Euler topology, cf. Eqs. (9-10) and the Discussion of the main text. We show the corresponding topologically-bounded gaps of Chern and Euler phases in Fig.~\ref{fig:S1}.

{\it (ii)} That translates into a completely different structure of the bound derivations. In the Euler case, the excitations from and to both of the Euler bands at the same time were considered, whereas in the Chern case, it is excitations from one (valence) Chern band to another (conduction) Chern band, see Fig.~\ref{fig:S1}. On removing a $\mathcal{T}$-breaking perturbation to connect to the Euler case, the Chern bound would need to reduce to transitions from lower to higher Euler band, which is not bounded by the invariant in the Euler case. Similarly in the optical weight bounds: the bound is on absorption (1) to and (2) from the Euler bands. Therefore, there is no evolution of the bound on removing the $\mathcal{C}_2\mathcal{T}$-breaking perturbation, and these bounds cannot be directly connected, as indicated in Fig.~\ref{fig:S1}.

{\it (iii)} Related to that, fundamentally one bound is in general not reducible onto another, as in one case auxiliary, purely mathematically-motivated quantum states, unlike the actual energy eigenstates corresponding to the Euler bands, were used to exploit the positive-semidefiniteness of the quantum-geometric tensor to derive the bound. In the other case, for Chern bands, actual eigenstates were used, without any complexification. In the Chern case, the objects in the bound -- quantum metric and Berry curvature -- both enter the response functions (optical and dc conductivities). On the other hand, for Euler bands, only bounds on the metric enter the optical responses (as the Berry curvature, and hence the circular dichroism, vanish), whereas the Euler curvature and the Euler invariant do not enter any previously known physical observables. This is in strong contrast with the bounds involving Berry curvature and the associated Chern invariant that enters the anomalous Hall conductivity and the TKNN invariant~\cite{PhysRevLett.49.405}, and this distinction also underpins the fact that the Euler invariant is more difficult to probe experimentally.

\section{Interband dc conductivity and magic-angle twisted bilayer graphene}\label{app:appE}

We start by noting that Eqs.~\eqref{eq:interbandDC} and~\eqref{opt_metric} together imply
\beq{eq:DCcon_met}
    \sigma_{\text{inter}}^{xx}(0) + \sigma_{\text{inter}}^{yy}(0) 
    = \frac{2 e^2}{\hbar} \sum_{m,n} \int \frac{\dd^2 k}{(2\pi)^2} f_{mn} (g^{mn}_{xx} + g^{mn}_{yy}) = \frac{2}{\pi}\Big(W^1_{xx}(\infty) + W^1_{yy}(\infty)\Big),
\eeq
which on insertion into Eq.~\eqref{eq:DCcon_met}, and the use of inequality in the main text, yields the bound, 
\beq{eq:DCcon}
    \sigma_{\text{inter},(\text{occ})}^{xx}(0) + \sigma_{\text{inter},(\text{occ})}^{yy}(0) +
    \sigma_{\text{inter},(\text{unocc})}^{xx}(0) + \sigma_{\text{inter},(\text{unocc})}^{yy}(0) \geq \frac{4e^2}{h} |\chi|.
\eeq  
This is the consequential result for Euler bands, arising from the bound on combined optical weight. In the case of Euler phases beyond the flat-band degenerate limit, the bound indeed holds within the independent particle picture, however, additional intraband contributions to the dc conductivity arise, %which can partly cancel the interband contribution on summation,% 
yielding a \textit{total} dc conductivity which manifestly does not satisfy the bound. In the real material setting embedded in the TBG context, one can fix occupation numbers to $f^{\chi}_{l\kv} = \frac{n+4}{8}$ in TBG (8 flat bands, in 4 pairs of bands with $\chi = 1$, for each spin and each valley), with the filling factor $n \in \small[ -4, 4 \small]$. We would obtain using Eq.~\eqref{eq:DCcon_met},

\beq{}
   \sigma_{(n)}^{aa}(0) \equiv \left(\frac{n+4}{8}\right) \sigma_{\text{inter},(\text{occ})}^{aa}(0) + \left(\frac{-n+4}{8}\right) \sigma_{\text{inter},(\text{unocc})}^{aa}(0).
\eeq

Then, on choosing without loss of generality: ${\sigma^{xx} \geq \sigma^{yy}}$, and accounting for the valley and spin degeneracies, we have:
\beq{eq:OGbound}
 \sigma_{(n)}^{xx}(0) + \sigma_{(-n)}^{xx}(0) \geq \frac{8e^2}{h} |\chi|,
\eeq
which can be contrasted with the experiments \cite{Kim_2017, Cao_2018}. Note that the finding of the bound on dc conductivity is subject to the range of applicability of the independent-particle picture \cite{huhtinen2022conductivity}. Namely, we observe that the dc conductivity of TBG is significantly smaller for most of the dopings $n$, which can be attributed to the strong correlations present in the material. Especially, at the charge neutrality point (${n=0}$), TBG is known to be a correlated insulator \cite{Lu_2019, Saito_2020}, strongly violating the bound. However, around ${n=\pm2}$ \cite{Cao_2018}, the bound is satisfied. 
%showing that the Euler topology could against the interactions under such doping. 
We stress that the effects violating the bound are beyond the independent particle picture, in which the Euler invariant is defined, consequences of which in terms of generalizations of quantum metric to interacting many-body systems, are beyond the scope of this work. Alternatively, the breakdown of the bound could be associated with the violation of the $\mathcal{C}_2\mathcal{T}$-symmetry, which might be even not preserved on average, as the real material is subject to the naturally present strains or disorder \cite{jankowski2023disorderinduced}, or when the symmetry is broken by the interactions. This insight further motivates pursuits studying quantum metric beyond the independent-particle picture, as well as its interplay with many-band invariants such as the Euler class, which is left for future work.

\section{Bounds on optical weight from Fermi's golden rule}\label{app:appF}

%{Transition rates}

In this section we elaborate on the possibility of probing Euler topology with optical transitions, as described by Fermi's golden rule (FGR). This is in close analogy to the deduction of Chern topology from magnetic circular dichroism \cite{doi:10.1126/sciadv.1701207}. 
When linearly-polarized light of frequency $\om$ and electric field amplitude $\mathcal{E}$ couples to the electrons with wavevector $\kv$ within a material, FGR give the rates of vertical transitions as
\beq{}
\begin{split}
    \Gamma_{x}(\om, \kv) = \frac{2\pi e^2\mathcal{E}^2 }{\hbar^2} \sum^{\text{occ}}_n \sum^{\text{unocc}}_m \left|\bra{\psi_{n\kv}} \hat{x} \ket{\psi_{m\kv}}\right|^2 f_{nm} \delta(E_m - E_n - \hbar \om),
\end{split}
\eeq
\beq{}
\begin{split}
    \Gamma_{y}(\om, \kv) = \frac{2\pi e^2\mathcal{E}^2 }{\hbar^2} \sum^{\text{occ}}_n \sum^{\text{unocc}}_m \left|\bra{\psi_{n\kv}} \hat{y} \ket{\psi_{m\kv}}\right|^2 f_{nm} \delta(E_m - E_n - \hbar \om),
\end{split}
\eeq
for light linearly-polarized in $x$ and $y$ correspondingly. Here, as in the previous sections, we denote the single-particle Bloch eigenstates as $\ket{\psi_{n\kv}} = e^{i\kv \cdot \rv}\ket{u_{n}}$, where we have suppressed the momentum label $\kv$ in $\ket{u_{n,\kv}}$ for simplicity. On rewriting the transition dipole matrix elements as the corresponding tangent vectors, we obtain (for $a = x,y$):
\beq{}
\begin{split}
    \Gamma_{a}(\om, \kv) = \frac{2\pi e^2\mathcal{E}^2 }{\hbar^2} \sum^{\text{occ}}_n \sum^{\text{unocc}}_m \left|\bra{u_n} \ii \partial_{a} \ket{u_m}\right|^2 f_{nm} \delta(E_m - E_n - \hbar \om),
\end{split}
\eeq
which can be further rewritten in terms of the non-Abelian Berry connection $A^a_{nm}$ and quantum metric $g^{nm}_{aa} = |A^a_{nm}|^2$ as
\beq{}
\begin{split}
    \Gamma_{a}(\om, \kv) = \frac{2\pi e^2\mathcal{E}^2 }{\hbar^2} \sum^{\text{occ}}_n \sum^{\text{unocc}}_m \left|A^a_{nm}\right|^2 f_{nm} \delta(E_m - E_n - \hbar \om) = \frac{2\pi e^2 \mathcal{E}^2 }{\hbar^2} \sum^{\text{occ}}_n \sum^{\text{unocc}}_m g^{nm}_{aa} f_{nm} \delta(E_m - E_n - \hbar \om) .
\end{split}
\eeq
Hence, we observe that the non-Abelian Berry connection elements can be reconstructed from the electric dipole (E1) transition rates induced by the linearly-polarized light, as given by FGR. In particular, on tuning the light frequency, it is possible to access the magnitudes of elements connecting different Euler bands ($|A^x_{12}|, |A^y_{12}|$), from which the Euler invariant can be reconstructed, as we discuss in the main text.

On the other hand, when circularly-polarized light of frequency $\om$ with electric field $\mathcal{E}$ couples to the electrons with wavevector $\kv$ within a material, FGR gives the rate of vertical transitions as
\beq{}
\begin{split}
    \Gamma_{\pm}(\om, \kv) = \frac{2\pi e^2\mathcal{E}^2 }{\hbar^2} \sum^{\text{occ}}_n \sum^{\text{unocc}}_m \left|\bra{u_n} \partial_{k_x} \pm \ii \partial_{k_y} \ket{u_m}\right|^2 f_{nm} \delta(E_m - E_n - \hbar \om),
\end{split}
\eeq
where `$+$' and `$-$' denote left- and right-circularly-polarized (LCP and RCP) light respectively. At zero temperature, this can be written
\beq{}
\begin{split}
    \Gamma_{\pm}(\om, \kv) = \frac{2\pi e^2\mathcal{E}^2 }{\hbar^2} \left(\sum_{mn} (r_{nm}^x r_{mn}^x + r_{mn}^y r_{nm}^y) \pm \ii\sum_{nm} {F}^{mn}_{xy}\right) \delta(E_m - E_n - \hbar \om)
\end{split}
\eec
where the elements $r^a_{mn}$ are defined in Eq. \ref{eq:ramn}. It follows that
\beq{}
\begin{split}
    \Gamma_{+}(\om, \kv) + \Gamma_{-}(\om, \kv) = \frac{2\pi e^2\mathcal{E}^2 }{\hbar^2} \sum_{mn} (g^{nm}_{xx} + g^{nm}_{yy}) \delta(E_m - E_n - \hbar \om)
\end{split}
\eec
so, by performing a double integral to find the total absorption rates $\Gamma^{\text{tot}}_{\pm} = \int \dd \om \int_{\text{BZ}} \frac{\dd^2k}{(2\pi)^2} \Gamma_{\pm}(\om, \kv)$, we find
\begin{equation}
\begin{split}
    \tilde{\Gamma}^\text{tot} = \tilde{\Gamma}^\text{tot}_{+} + \tilde{\Gamma}^\text{tot}_{-} = \frac{2\pi e^2\mathcal{E}^2 }{\hbar^2} \int \dd^2 k~ \Tr\textbf{g}~ \geq \frac{2\pi e^2 \mathcal{E}^2 }{\hbar^2} |\chi|,
\end{split}
\end{equation}
where for each of the rates $\Gamma^{\text{tot}}_{\pm}$ we took a combination $\tilde{\Gamma}^{\text{tot}}_{\pm} \equiv \Gamma^{\text{tot}}_{\pm,\text{unocc}} + \Gamma^{\text{tot}}_{\pm,\text{occ}}$. Here, the notation `occ' and `unocc' represents the configurations discussed in Sec.~\ref{app:appC}.
Once again, the result uses the inequality shown in Ref.~\cite{PhysRevLett.124.167002, bouhon2023quantum}, and is the FGR manifestation of the bound on optical weights, given the transition rates due to the linearly-polarized light are described by the average/total captured by the sum: $\Gamma^\text{tot} = \Gamma^{\text{tot}}_{+} + \Gamma^{\text{tot}}_{-}$.

We note that in Euler phases, where the elements of the Abelian Berry curvature ${F}^{mn}_{xy} = 0$ and the elements $r_{nm}^i r_{mn}^i$, with $i,j = x,y$, are real, the difference of the rates $\Delta \Gamma = \Gamma^{\text{tot}}_{+} - \Gamma^{\text{tot}}_{-} = 0$. Hence, despite the lower bound on the absorption of the combined circularly-polarized-lights, Euler phases show no circular dichroism. This may be contrasted with Chern insulators, where indeed the quantized magnetic circular dichroism follows from the momentum-resolved transition rates. We recall that this may be shown as follows: writing 
\beq{}
\begin{split}
    \Delta \Gamma(\kv) = \Gamma_{+}(\kv) - \Gamma_{-}(\kv)
    = \ii\frac{4\pi e^2\mathcal{E}^2}{\hbar^2}\sum^{\text{occ}}_n \sum^{\text{unocc}}_m \Bigg(\bra{u_n} \ket{\partial_y u_m}  \bra{u_m} \ket{\partial_x u_n} - \bra{u_n} \ket{\partial_x u_m}  \bra{u_m} \ket{\partial_y u_n} \Bigg)
\end{split}
\eec
and recognizing using the quantum geometric tensor $\mathcal{F}_{n} = -2 \sum^{\text{unocc}}_m \Im (r^x_{nm} r^y_{nm}) $ we have
\beq{}
\begin{split}
    \Delta \Gamma(\kv)
    = \frac{4\pi e^2\mathcal{E}^2}{\hbar^2}\sum^{\text{occ}}_n \mathcal{F}_{n},
\end{split}
\eeq
and we may perform the integral over the BZ, $ \Delta \Gamma = \int_{\text{BZ}} \frac{\dd^2k}{(2\pi)^2} \Delta \Gamma(\kv) $ to find \cite{doi:10.1126/sciadv.1701207}:
\beq{}
\begin{split}
    \Delta \Gamma
    = \frac{2 e^2\mathcal{E}^2}{\hbar^2}\sum^{\text{occ}}_n C_n
\end{split}
\eec
where $C_n = \frac{1}{2\pi} \int_{\text{BZ}} \dd^2k~\mathcal{F}_{n}$ is the Chern number of the $n^{\text{th}}$ band.

Finally, we may also consider the use of second- or higher-order processes captured by FGR to measure the Euler curvature. Consider a transition from the lower Euler band, which we label `1', to a higher unoccupied band, and then back to the upper Euler band, labelled `2'. The corresponding matrix element is of the form
\beq{}
    \mathcal{M}^{(2)}_{\text{if}} \propto \frac{\bra{u_1} \partial_x \pm \ii \partial_y \ket{u_m} \bra{u_m} \partial_x \pm \ii \partial_y \ket{u_2} }{E_2 - E_m + \hbar\om}.
\eeq
If we restrict to 3 bands and insert a resolution of identity, we have
\beq{}
    \mathcal{M}^{(2)}_{\text{if}} \propto \frac{(r^x_{13} \pm \ii r^y_{13}) (r^x_{32} \pm \ii r^y_{32})  }{E_2 - E_3 + \hbar\om}.
\eeq
Hence, for mixed transitions, where the first perturbation is due to LCP and the second to RCP, we have
\beq{}
    \mathcal{M}^{(2)}_{\text{if}} \propto \frac{\pm \ii \text{Eu} + r^y_{13} r^y_{32} + r^x_{13} r^x_{32} }{E_2 - E_3 + \hbar\om}.
\eeq
In general, the terms appearing after the Euler curvature in this expression are non-vanishing, which means that the Euler curvature cannot be directly extracted from this matrix element. However, by forming a difference of matrix elements, by perturbing with LCP, and then with RCP vs.~the other way around, we may write
\beq{}
    \mathcal{M}^{+(2)}_{\text{if}} - \mathcal{M}^{-(2)}_{\text{if}} \propto  \ii \frac{ 2\text{Eu} }{E_2 - E_3 + \hbar\om},
\eeq
which could, at least in principle, be measured by obtaining the absolute values of matrix elements in momentum-resolved experiments.

Similarly, another matrix element that is directly related to the Euler curvature is that corresponding to the response to two-photon excitations with  \textit{linearly} polarized light: if the first part of the virtual transition is due to polarization along the $y$-direction, and the second to polarization along the $x$-direction, then the difference between this matrix element and its reversed counterpart is
\begin{equation}
     \mathcal{M}^{yx(2)}_{\text{if}} - \mathcal{M}^{xy(2)}_{\text{if}} \propto \frac{ 2\text{Eu} }{E_2 - E_3 + \hbar\om} = \frac{\bra{u_1}  \ket{\partial_y u_3} \bra{u_3}  \ket{ \partial_x  u_2} - \bra{u_2} \ket{\partial_y u_3} \bra{u_3}  \ket{ \partial_x  u_1}}{E_2 - E_3 + \hbar\om}.
\end{equation}
Measuring such second-order matrix elements in  a momentum-, as well as frequency-resolved way could pose a significant experimental challenge. However, as we show in the main text, it is not necessary to extract the curvature experimentally from these types of second-order transitions, as it can be directly reproduced from the vorticity of the connection $\vec{A}_{1,2}$, which itself can be probed with the first-order (E1) transitions (given sufficient momentum-space resolution). 

\section{Analytical results for the continuum model of Euler nodes}\label{app:appG}

Following the introduction of the effective model for an Euler node in the main text, we obtain a range of analytical results for the optical properties of this system. We begin by computing the non-Abelian Berry connection: the diagonal elements $\vec{A}_{1,1}(\kv)=\vec{A}_{2,2}(\kv)=0$ vanish, while the off-diagonal terms are
\begin{equation}
    \vec{A}_{1,2}(\kv)=-\vec{A}_{2,1}(\kv)=\frac{\chi}{k^2}(-k_y, k_x).
\end{equation}
From this we find the quantum metric elements $g^{12}_{ab}$ (Sec.~\ref{app:appA}):
\begin{subequations}
    \begin{align}
    g^{12}_{xx} =&{} -A^x_{1,2} A^x_{2,1} = \frac{\chi^2 k^2_y}{k^4},\\
    g^{12}_{xx} =&{} -A^y_{1,2} A^y_{2,1} = \frac{\chi^2 k^2_x}{k^4},\\
    g^{12}_{xy} =&{} g^{12}_{yx} = -A^x_{1,2} A^y_{2,1} = -\frac{\chi^2 k_x k_y}{k^4}.
    \end{align}
\end{subequations} 
For incident radiation of frequency $\omega$, the delta function that appears in a number of the optical response formulae in Sec.~\ref{app:appB} may be expressed as
\begin{equation}\label{eq:DiraCDelta}
   \delta(\om_{nm}(\kv)-\omega) = \oint_{\om_{nm}(\kv')=\omega}
\dd k'\,\frac{\delta(\kv-\kv')}{|\nabla_{\kv'}\om_{nm}(\kv')|},
\end{equation}
where for the introduced rotationally-symmetric two-band model $\om_{12}(\kv) = 2\alpha |k|^{2\chi}$, with a constant $\alpha$ (see Fig.~1 in the main text). The region where transitions at frequency $\om$ can occur is a circle of radius $k'=(\omega/2\alpha)^{\frac{1}{2\chi}}$. It follows that $|\nabla_{\kv}\om_{12}|=|2\alpha \nabla_{\kv}k^{2\chi}|=4 \alpha \chi k^{2\chi-1}$. Using Eq. \ref{eq:AC_OptCond} for the optical conductivity, we have 
\beq{}
\begin{split}
    \sigma^{ab}(\omega) = \frac{ \om e^2}{4\pi \hbar} \int \frac{\dd^2 k}{(2\pi)^2}~ (f_{1\kv} - f_{2\kv}) g^{12}_{ab} \delta(\om_{12} - \omega)
\end{split}
\eeq
where at half-filling and in the zero-temperature limit $f_{2\kv} = 0$ and $f_{1\kv} = 1$ (of course, the occupations may be altered by doping the chemical potential away from the node). Through the use of the identity Eq. \ref{eq:DiraCDelta} this becomes
\beq{}
\begin{split}
    \sigma^{ab}(\omega) = \frac{ \om e^2}{4\pi \hbar} \oint_{\om_{nm}=\omega}
\dd \theta~ |k'| \frac{g^{12}_{ab}}{4 \alpha \chi |k'|^{2\chi-1}}.
\end{split}
\eeq
By writing $k_x = |k| \cos \theta$, $k_y = |k| \sin \theta$, we then obtain
\beq{}
\begin{split}
    \sigma^{xx}(\omega) = \frac{ \om e^2}{4\pi \hbar} \int^{2\pi}_0
\dd \theta~ \frac{\chi \cos^2 \theta}{4 \alpha |k'|^{2\chi}} = \frac{e^2}{8 \hbar} |\chi|,
\end{split}
\eeq
with $\sigma^{xx}(\omega) = \sigma^{yy}(\omega)$ and $\sigma^{xy}(\omega) = 0$.

At second-order, one finds the linear injection photoconductivity
\beq{}
    \sigma^{cab}_{\text{inj},L}(\om) = -\frac{ e^3 \tau}{4\pi \hbar^2} \int^{2\pi}_0 \dd \theta~ |k'| \frac{g^{12}_{ab}}{4 \alpha \chi |k'|^{2\chi-1}} \partial_c \om_{mn},
\eeq
and the circular shift photoconductivity
\beq{}
    \sigma^{cab}_{\text{shift},C}(\om) = i\frac{ e^3}{8 \pi \hbar^2} \int^{2\pi}_0 \dd \theta~ \frac{|k'|}{4 \alpha \chi |k'|^{2\chi-1}} (A^{a}_{12} \partial_c A^{b}_{21} - A^{b}_{12} \partial_c A^{a}_{21}).
\eeq
The third-order jerk photoconductivity is \cite{PhysRevLett.121.176604}
\beq{}
    \sigma^{cdab}_{\text{jerk}}(\om) = -\frac{ e^4 \tau^2}{2\pi \hbar^3} \int^{2\pi}_0 \dd \theta~ |k'| \frac{g^{12}_{ab}}{4 \alpha \chi |k'|^{2\chi-1}} \partial_c \partial_d \om_{mn}.
\eeq
We omit the third-order shift photoconductivities; they may be computed in a similar manner. For the injection conductivities
we have
\begin{subequations}
    \begin{align}
    \sigma^{xxx}_{\text{inj},L}(\om) ={}& -\frac{ e^3 \tau}{4\pi \hbar^2} \int^{2\pi}_0 \dd \theta~  \frac{\chi \cos^2 \theta}{4 \alpha |k'|^{2\chi}} 4 \alpha \chi |k'|^{2\chi-1} \cos{\theta} = -\frac{\pi e^3 \tau}{\hbar^2} \int^{2\pi}_0 \dd \theta~  \frac{\chi^2 \cos^3 \theta}{|k'|} =0,\\
    \sigma^{yxx}_{\text{inj},L}(\om) ={}& -\frac{ e^3 \tau}{4\pi \hbar^2} \int^{2\pi}_0 \dd \theta~  \frac{\chi \cos^2 \theta}{4 \alpha |k'|^{2\chi}} 4 \alpha \chi |k'|^{2\chi-1} \cos{\theta} = -\frac{\pi e^3 \tau}{\hbar^2} \int^{2\pi}_0 \dd \theta~  \frac{\chi^2 \cos^2 \theta \sin \theta}{|k'|} =0.
    \end{align}
\end{subequations}
Similarly, the circular shift currents vanish, which is a general result of the fact that second-order responses always vanish in centrosymmetric systems, as is the case for the rotationally symmetric node described by our model. This motivates the study of the higher-order jerk currents:
\begin{subequations}
    \begin{align}
    \sigma^{xxxx}_{\text{jerk}}(\om) ={}& -\frac{ e^4 \tau^2}{2\pi \hbar^3} \int^{2\pi}_0 \dd \theta~  \frac{\chi \cos^2 \theta}{4 \alpha |k'|^{2\chi}} 4 \alpha \chi \left( (2\chi - 1) \cos^2 \theta  + \sin^2 \theta\right) |k'|^{2\chi-2},\\
    \sigma^{xxyy}_{\text{jerk}}(\om) ={}& -\frac{ e^4 \tau^2}{2\pi \hbar^3} \int^{2\pi}_0 \dd \theta~  \frac{\chi \cos^2 \theta}{4 \alpha |k'|^{2\chi}} 4 \alpha \chi \left( (2\chi - 1) \sin^2 \theta  + \cos^2 \theta\right) |k'|^{2\chi-2},\\
    \sigma^{xyxy}_{\text{jerk}}(\om) ={}& -\frac{ e^4 \tau^2}{2\pi \hbar^3} \int^{2\pi}_0 \dd \theta~  \frac{-\chi \sin \theta \cos \theta}{4 \alpha |k'|^{2\chi}} 8 \alpha \chi (\chi - 1) |k'|^{2\chi-2} \sin{\theta} \cos{\theta},\\
    \sigma^{xxxy}_{\text{jerk}}(\om) ={}& -\frac{ e^4 \tau^2}{2\pi \hbar^3} \int^{2\pi}_0 \dd \theta~  \frac{\chi \cos^2 \theta}{4 \alpha |k'|^{2\chi}} 8 \alpha \chi (\chi - 1) |k'|^{2\chi-2} \sin{\theta} \cos{\theta},
    \end{align}
\end{subequations}
where, notably, only terms with angular integrals $\int^{2\pi}_0 \dd \theta~ \sin^2{\theta} \cos^2{\theta} = \frac{\pi}{4}$,  $\int^{2\pi}_0 \dd \theta~ \cos^4{\theta} = \int^{2\pi}_0 \dd \theta~ \sin^4{\theta} = \frac{3\pi}{4}$ survive. Hence,
\begin{subequations}
    \begin{align}
    \sigma^{xxxx}_{\text{jerk}}(\om) ={}& \sigma^{yyyy}_{\text{jerk}}(\om) = -\frac{ e^4 \tau^2}{4 \hbar^3} \chi^2 (3\chi - 1) { \left( \frac{2\alpha}{\om} \right)^{\frac{1}{\chi}}},\\
    \sigma^{xxyy}_{\text{jerk}}(\om) ={}& \sigma^{yyxx}_{\text{jerk}}(\om) = -\frac{ e^4 \tau^2}{4 \hbar^3} \chi^2 (\chi + 1) { \left( \frac{2\alpha}{\om} \right)^{\frac{1}{\chi}}},\\
    \sigma^{xyxy}_{\text{jerk}}(\om) ={}&\sigma^{yxyx}_{\text{jerk}}(\om) = \sigma^{xyyx}_{\text{jerk}}(\om) = \sigma^{yxxy}_{\text{jerk}}(\om) = \frac{ e^4 \tau^2}{4 \hbar^3} \chi^2 (\chi - 1) { \left( \frac{2\alpha}{\om} \right)^{\frac{1}{\chi}}},
    \end{align}
\end{subequations}
which, as expected, are model-dependent ($\alpha$ enters through dispersion relation), and vary with system properties such as the lattice parameter. However, \textit{ratios} of jerk conductivities, as indicated in the main text, are nevertheless independent of the particular value of $\alpha$.

\clearpage

\section*{Supplementary References}
\bibliographystyle{apsrev4-1}
\bibliography{supp_references.bib}